\documentclass[sigconf]{acmart}
\usepackage{tikz}
\usepackage{xcolor}

\newcommand*\circled[1]{\tikz[baseline=(char.base)]{
            \node[shape=circle,draw,inner sep=0.5pt] (char) {#1};}}

\AtBeginDocument{%
  \providecommand\BibTeX{{%
    \normalfont B\kern-0.5em{\scshape i\kern-0.25em b}\kern-0.8em\TeX}}}

\copyrightyear{2021} 
\acmYear{2021} 
\setcopyright{acmcopyright}\acmConference[CHI '21]{CHI Conference on Human Factors in Computing Systems}{May 8--13, 2021}{Yokohama, Japan}
\acmBooktitle{CHI Conference on Human Factors in Computing Systems (CHI '21), May 8--13, 2021, Yokohama, Japan}
\acmPrice{15.00}
\acmDOI{10.1145/3411764.3445444}
\acmISBN{978-1-4503-8096-6/21/05}

\begin{document}

%%
%% The "title" command has an optional parameter,
%% allowing the author to define a "short title" to be used in page headers.
\title{Causal Perception in Question-Answering Systems}

%%
%% The "author" command and its associated commands are used to define
%% the authors and their affiliations.
%% Of note is the shared affiliation of the first two authors, and the
%% "authornote" and "authornotemark" commands
%% used to denote shared contribution to the research.

\author{Po-Ming Law}
\affiliation{%
  \institution{Georgia Institute of Technology}
}
\email{pmlaw@gatech.edu}

\author{Leo Yu-Ho Lo}
\affiliation{%
  \institution{The Hong Kong University of Science and Technology}
}
\email{yhload@cse.ust.hk}

\author{Alex Endert}
\affiliation{%
  \institution{Georgia Institute of Technology}
}
\email{endert@gatech.edu}

\author{John Stasko}
\affiliation{%
  \institution{Georgia Institute of Technology}
}
\email{stasko@cc.gatech.edu}

\author{Huamin Qu}
\affiliation{%
  \institution{The Hong Kong University of Science and Technology}
}
\email{huamin@cse.ust.hk}

%%
%% By default, the full list of authors will be used in the page
%% headers. Often, this list is too long, and will overlap
%% other information printed in the page headers. This command allows
%% the author to define a more concise list
%% of authors' names for this purpose.
\renewcommand{\shortauthors}{Law, Lo, Endert, Stasko, and Qu}

%%
%% The abstract is a short summary of the work to be presented in the
%% article.
\begin{abstract}

Root cause analysis is a common data analysis task. While question-answering systems enable people to easily articulate a why question (e.g., why students in Massachusetts have high ACT Math scores on average) and obtain an answer, these systems often produce questionable causal claims. To investigate how such claims might mislead users, we conducted two crowdsourced experiments to study the impact of showing different information on user perceptions of a question-answering system. We found that in a system that occasionally provided unreasonable responses, showing a scatterplot increased the plausibility of unreasonable causal claims. Also, simply warning participants that correlation is not causation seemed to lead participants to accept reasonable causal claims more cautiously. We observed a strong tendency among participants to associate correlation with causation. Yet, the warning appeared to reduce the tendency. Grounded in the findings, we propose ways to reduce the illusion of causality when using question-answering systems.

\end{abstract}

%%
%% The code below is generated by the tool at http://dl.acm.org/ccs.cfm.
%% Please copy and paste the code instead of the example below.
%%


%%
%% Keywords. The author(s) should pick words that accurately describe
%% the work being presented. Separate the keywords with commas.

%%
%% This command processes the author and affiliation and title
%% information and builds the first part of the formatted document.
\maketitle

\section{Introduction}

Root cause analysis is a common task during data analysis. Such analysis provides explanations for events in business processes, observations about human behaviours, and phenomena in society. A business analyst, for instance, may seek explanations for a revenue decrease to identify supply chain bottlenecks and marketing strategies~\cite{business}. To help people acquire this important skill, colleges and online learning platforms have offered courses on root cause analysis~\cite{rca1, rca2}.

The need for root cause analysis skills is not only limited to professional analysts. Open data create opportunities for anyone to engage in personal data projects. Visualization hobbyists, for example, may conduct data analysis on public data and create fascinating visualizations on platforms such as Makeover Monday~\cite{makeover}. Individual citizens might analyze data about the social issues they are concerned about and write a blog post about the analysis~\cite{medium1, medium2}. However, root cause analysis could be challenging to these people since they might lack domain knowledge and analysis skills.

Systems with question-answering functionality present a resource that people can utilize to explain data observations even without significant expertise in data analysis. Users of these systems can easily articulate their why questions through natural language~\cite{analyza} or point and click~\cite{explainData}. The systems then employ advanced statistical analysis to infer answers. Some technologists believe that question-answering interfaces will become the norm in analytics platforms~\cite{NLOpinion}.

\begin{figure}[t!]
	\centering
	\includegraphics[width=\linewidth]{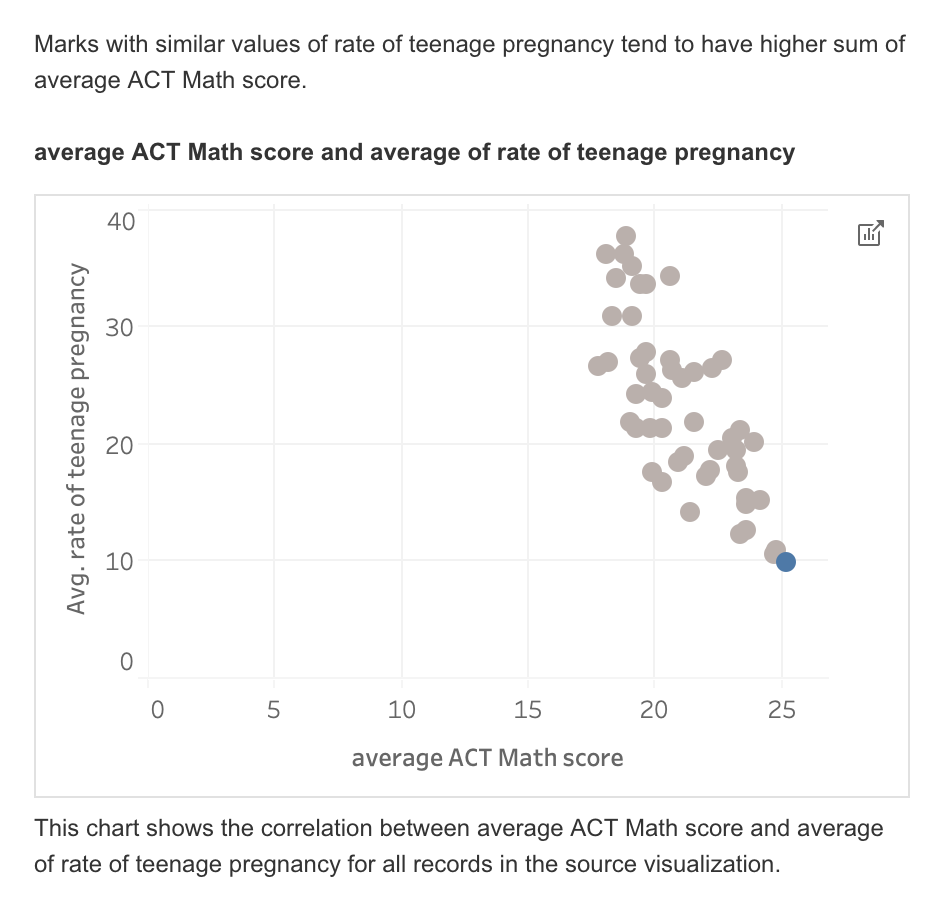}
	\caption{An answer generated by Tableau Explain Data~\cite{explainData}. The user asks about the high ACT Math score in Massachusetts. Explain Data infers that teenage pregnancy rate and ACT Math score are negatively correlated and the low teenage pregnancy rate in Massachusetts might cause the high ACT Math score. It shows the data using a scatterplot in which each dot is a state and the blue dot is Massachusetts.}
	\Description{A screenshot of an explanation provided by Tableau Explain Data. It shows a scatterplot with average ACT Math score on the X axis and average rate of teenage pregnancy on the Y axis. Massachusetts is at the bottom right of the point cloud, indicating that it has a high average ACT Math score and a low average rate of teenage pregnancy. At the bottom of the scatterplot, there is a textual description, saying that ``This chart shows the correlation between average ACT Math score and average rate of teenage pregnancy for all records in the source visualization.''}
	\label{explain}
\end{figure}

However, causal inference from observational data (as opposed to randomized experiments) is challenging~\cite{causalInference}. These question-answering systems often produce unintuitive answers to a user's why questions. Figure~\ref{explain} shows Explain Data, a question-answering functionality in Tableau~\cite{explainData}. The user observes that Massachusetts has the highest average ACT Math score among all US states. Being curious, she asks Explain Data to provide explanations for the high score. Explain Data infers that the rate of teenage pregnancy is negatively correlated with ACT Math score and that the low rate of teenage pregnancy in Massachusetts may lead to the high ACT Math score. The veracity of the explanation is questionable.

When these systems do not always provide reliable results, a concern is their potential power to persuade people into believing causal claims (e.g., low teenage pregnancy rate in Massachusetts may lead to the high average ACT Math score) that may not be true. Does visualizing correlation (e.g., using a scatterplot) increase the plausibility of a causal claim and user trust in the system even when the claim does not make sense? Does warning users about the potential flaws in the system help them adopt the answers more cautiously? Answering these questions could help understand designs that ensure judicious use of the computational outputs.

This paper investigates the impacts of different information (a scatterplot, a description about correlation, and a warning message) shown alongside a causal claim on the perceptions of a question-answering system. We conducted two crowdsourced studies with 200 participants each. In both studies, participants reviewed a series of answers to why questions. These answers were presented with different designs. Across different designs, we compared the perceived plausibility of the causal claims, user trust in the question-answering system, the awareness of the system's flaws, and users' tendency to associate correlation with causation. Whereas the first study presented answers with different degrees of plausibility, the second study presented only reasonable answers.

From the first study, we found that participants tended to disagree less with an unreasonable causal claim when a scatterplot was presented alongside the claim. In contrast, participants appeared to accept a reasonable causal claim more cautiously when they were shown a simple warning about the system's potential confusion of correlation and causation. We further observed a general tendency among participants to associate correlation with causation, but the warning seemed to reduce the tendency. We did not observe these effects in the second study where the system only provided reasonable causal claims. 

Question-answering systems often employ data visualizations to provide context for their answers~\cite{visualConverse, vagueMod}. Our results reveal that these systems could leverage the persuasive power of visualizations to create an illusion of causality: Although scatterplots only provide evidence about correlation, presenting scatterplots next to a causal claim could increase users' tendency to agree with the claim. Based on our findings, we suggest that users should be skeptical when considering answers that are automatically generated and propose design ideas to encourage skepticism.

\section{Related Work}

Our work intends to understand how the visual design of answers to why questions might influence the perceptions of a question-answering system. We draw on research relating to the impact of visualization design on data interpretation as well as question-answering systems more broadly.

\subsection{Impact of Visualization Design on Data Interpretation}

Visualization design holds significant power to shape data interpretation~\cite{persuasive}. Researchers have investigated a wide range of factors such as knowledge, perceptual biases, and cognitive biases that influence the messages communicated to viewers.

Knowledge external to visualizations often affects how we interpret the visualizations. As users look at a visualization, they often apply their domain knowledge~\cite{knowledge}. Xiong et al.~\cite{curse} showed that this prior knowledge could prime a viewer to obtain a particular message from a visualization and lead the viewer to believe that other viewers would receive the same message. Besides prior knowledge, social information also affects data interpretation. Kim et al.~\cite{otherEyes} found that seeing others' expectations about the data influenced people's trust in the accuracy of the data.

Moreover, perceptual biases play a role in manipulating data interpretation~\cite{blackhat, deceptive}. For example, distorting the aspect ratio of a line chart can lead to an inaccurate assessment of trends in the data~\cite{lineChart}; truncating the y-axis in a bar chart exaggerates effect sizes~\cite{truncate}; the neighborhood of a bar in a bar chart can change the perceptions of the bar's height~\cite{neighborhood}. However, these biases could be mitigated through judicious design. For instance, Ritchie et al.~\cite{lie} showed that an animated transition from an untruncated bar chart to a truncated one could avoid misinterpretation.

Cognitive biases can further change the lens through which we interpret visualizations~\cite{cognitive}. An example is priming and anchoring effects. Calero Valdez et al.~\cite{priming} conducted experiments to show that the judgment of class separability in scatterplots depended on the scatterplots users saw before. Biases in data interpretation can also have consequences on decision making. Dimara et al.~\cite{attraction} provided evidence that the presence of dominated data points in a scatterplot influenced the judgement of which points were dominating.

Besides knowledge and biases, subtle design choices also matter to data interpretation: Titles can have a misleading impact on visualization interpretation~\cite{memorability, rhetoric, frame1, frame2}; visual embellishments can affect the insights we gain from visualizations~\cite{icons, styles}.

While correlation does not imply causation, it is easy to confuse them, leading to an illusion of causality~\cite{causalIllusion}. Xiong et al.'s~\cite{correlation} found that this illusion would increase with the aggregation level of data visualizations. Instead of studying data aggregation, we investigated the effects of different information on perceived causality when using question-answering systems. Specifically, we studied whether two forms of correlational evidence (scatterplot and textual description about correlation) could create causal illusion and whether a simple warning could reduce the illusion.

\subsection{Question-Answering Systems and User Perception}

Technologists have developed question-answering systems to meet users' information needs in various domains including sports~\cite{sports}, work settings~\cite{work}, and data science~\cite{dataSci}. These systems exhibit a wide variety of designs. Some (e.g., conversational agents or chatbots) mimic natural human conversations and can understand a rich diversity of topics~\cite{siri}. Others resemble web search and focus only on a small set of tasks~\cite{me2, me1}.

In data visualization, researchers have developed natural language interfaces to facilitate visual data analysis~\cite{eviza, orko, flowsense}. Many of these systems aim to address specific usability challenges as users employ natural language for data analysis. For example, users' utterances are often ambiguous. Datatone utilizes ambiguity widgets to expose the ambiguity and let users correct the system's decisions~\cite{datatone}. Moreover, conversations happen in some context on which utterance semantics depend~\cite{evizeon}. To address this, Evizeon provides pragmatics support to retain contextual information and infer a user's meaning based on the context~\cite{evizeon}.

Another line of research focuses on understanding the impact of system behaviors and information presentation on the perceptions of these systems. Liao et al.~\cite{social} investigated how agent sociability influences user interactions with conversational agents. Ashktorab et al.~\cite{breakdown} studied preferences for different strategies to handle conversational breakdowns. Hearst et al.~\cite{visualConverse, vagueMod} investigated the visual designs of answers provided by a natural language interface and how users perceive these designs. In a similar vein, we intend to provide insights into how the visual design of answers to why questions might affect user perceptions of a question-answer system.

\section{Pre-Study: Collecting Causal Statements}

As a starting point to understand the appropriate presentation of answers to why questions, we focus on why questions about extremum (i.e., an extreme value). An example is why students in Massachusetts have high ACT Math scores on average (Fig.~\ref{explain}). Finding extremum is a common task during data analysis~\cite{lowLevel}. Also, functionality to answer such questions has emerged in commercial systems such as Tableau~\cite{explainData}. Findings from our studies could offer design guidelines in practice. 

In study 1, we showed participants a series of answers to why questions. We created answers with different visual designs and assessed user perceptions of the system given the designs. Due to the inherent challenges in causal inference~\cite{causalInference}, question-answering systems occasionally provide unreasonable answers to why questions. To emulate these systems, we selected causal claims with different levels of plausibility as answers presented to participants. To select these causal claims, we conducted a pre-study.

\subsection{Methods}

\subsubsection{Datasets}

We planned to generate causal claims that were backed up by observational data and considered using synthetic data. However, our goal was to study user perceptions of a system, and the credibility of the data might affect user perceptions. To control for the potential experimental confounds, we used real-world data instead.

We first curated a dataset about states in the US from sources including US Census Bureau~\cite{census}, National Center for Education Statistics~\cite{nces}, and Kaiser Family Foundation~\cite{kff}. The curated dataset has 258 attributes about demographics, healthcare, and education for each US state. We chose these topics because they are accessible to laypeople. This enabled participants to judge the plausibility of the generated causal claims based on common sense.

With the curated data table, we computed the Pearson correlation for all attribute pairs. To find attribute pairs with a potential causal relationship, we collected the ones with a high correlation (above 0.7 or below -0.7). For each of the 1522 attribute pairs with a high correlation (e.g., employment rate and poverty rate), we found a state (e.g., Mississippi) that has an extreme value for \textit{both} attributes and omitted the attribute pairs where such a state did not exist.

\begin{figure}[t!]
	\centering
	\includegraphics[width=\linewidth]{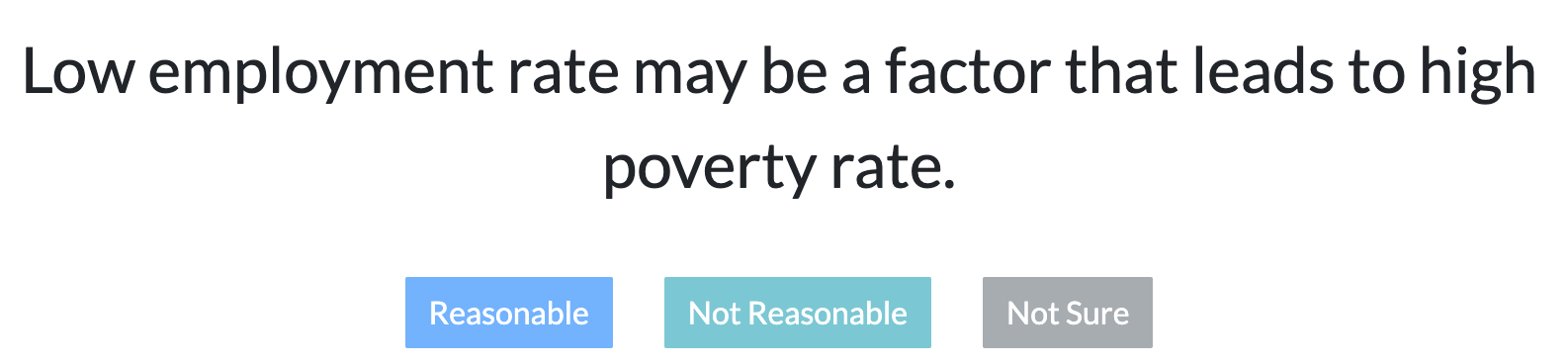}
	\caption{Interface used in the pre-study.}
	\Description{The interface participants interact with in the pre-study. It shows a causal claim ``Low employment rate may be a factor that leads to high poverty rate.'' Below the claim, there are three buttons with the labels ``Reasonable,'' ``Unreasonable,'' ``Not Sure.''}
	\label{prestudy-interface}
\end{figure}

Based on the collected attribute pairs and states, we generated causal claims (e.g., low employment rate in Mississippi may be a factor that leads to the high poverty rate in Mississippi). The plausibility of this claim can be affected by the plausibility of the causal relationship (e.g., employment rate affects poverty rate) and that of the information about the state (e.g., Mississippi has a high poverty rate). Since we intended to assess the plausibility of the causal relationship, we removed the states from the causal claims (see Fig.~\ref{prestudy-interface}).

An author carefully picked 30 reasonable claims, 30 unreasonable claims and 30 claims that were hard to tell if they were reasonable (hereafter, \textit{hard-to-tell claims}). We verified and ranked the plausibility of these claims through a study on Amazon Mechanical Turk (MTurk).

\begin{figure*}[t!]
	\centering
	\includegraphics[width=\linewidth]{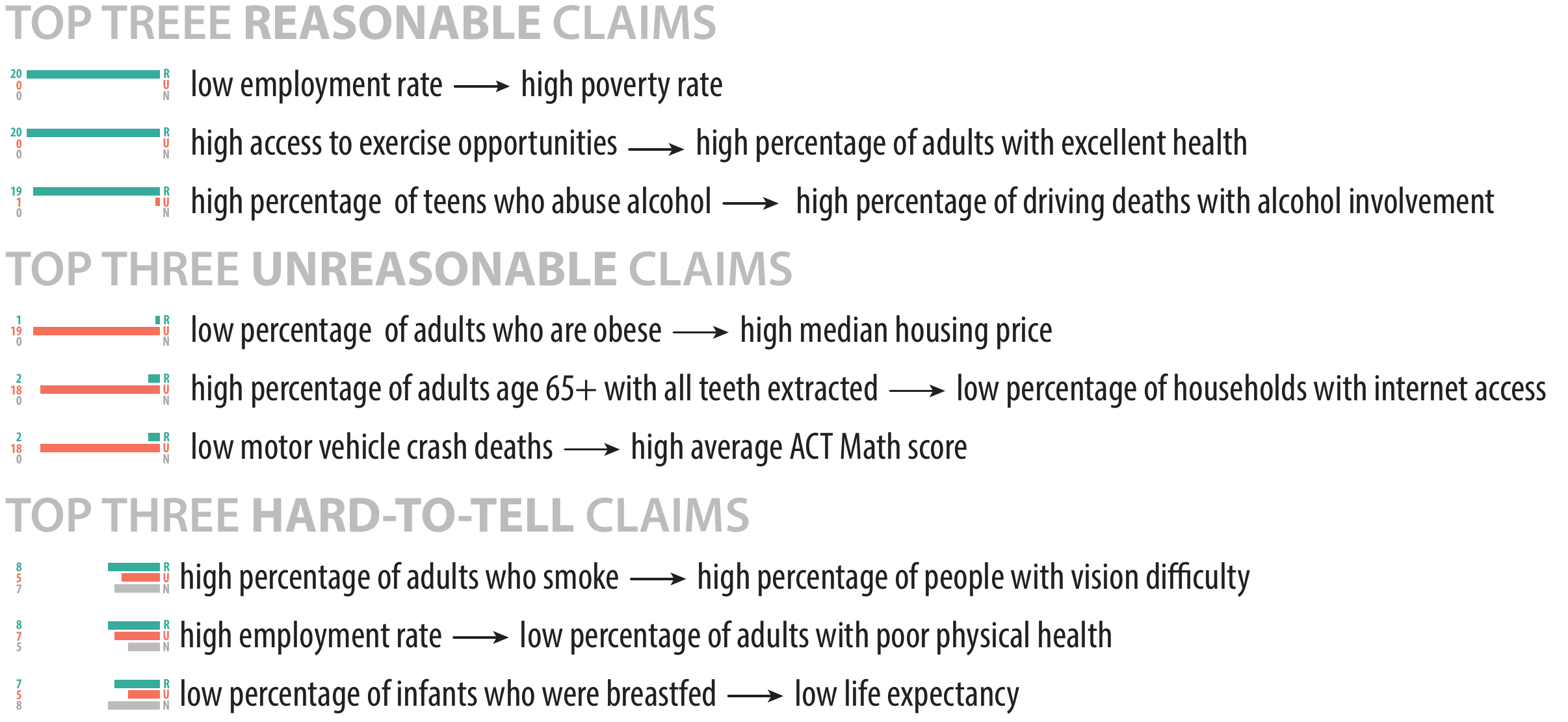}
	\caption{The top three reasonable, unreasonable, and hard-to-tell causal claims. Each row shows a causal claim (right) and a bar chart that visualizes the votes (left). The green, red, and gray bars represent the votes for \textit{Reasonable}, \textit{Unreasonable}, and \textit{Not Sure} respectively. For example, the most reasonable claim was ``a low employment rate may be a factor that leads to a high poverty rate.'' It got 20/20 votes for \textit{Reasonable} (R), 0/20 vote for \textit{Unreasonable} (U), and 0/20 vote for \textit{Not Sure} (N).}
	\Description{A ranked list of the top three reasonable claims, a ranked list of the top three unreasonable claims, and a ranked list of the top three hard-to-tell claims. The top three reasonable claims are ``a low employment rate may be a factor that leads to a high poverty rate,'' ``high access to exercise opportunities may be a factor that leads to a high percentage of adults with excellent health,'' and ``a high percentage of teens who abuse alcohol may be a factor that leads to a high percentage of driving deaths with alcohol involvement.'' The top three unreasonable claims are ``a low percentage of adults who are obese may be a factor that leads to a high median housing price,'' ``a high percentage of adults age 65+ with all teeth extracted may be a factor that leads to a low percentage of households with internet access,'' and ``low motor vehicle crash deaths may be a factor that leads to a high average ACT Math score.'' The top three hard-to-tell claims are ``a high percentage of adults who smoke may be a factor that leads to a high percentage of people with vision difficulty,'' ``a high employment rate may be a factor that leads to a low percentage of adults with poor physical health,'' and ``a low percentage of infants who were breastfed may be a factor that leads to a low life expectancy.''}
	\label{topClaims}
\end{figure*}

\subsubsection{Participants}

We randomly segmented the 90 claims into five batches of 18 claims and recruited 20 workers on MTurk to rate each batch (100 unique workers in total). We limited the tasks to workers in the United States and had an acceptance rate of 95\% or above. During data analysis, we omitted participants who failed to pass attention checks (but compensated them for participation). We recruited participants until reaching the target sample size for each batch. Participants were compensated \$1 for the study that took approximately 5-10 minutes.

Among the 100 participants, 55 were male, and 45 were female. They aged 22-64 (\textit{M}=35.5, \textit{SD}=11.2). Participants reported their educational attainment to be high school (8 participants), professional school (18), college (49), graduate school (17), PhD (7), and postdoctoral (1).

\subsubsection{Procedure}

Each participant was randomly assigned to rate one of the five batches of 18 claims. Participants first filled out a demographic survey on their gender, age, and highest education level. They then saw a series of 18 causal claims that were presented on separate pages (Fig.~\ref{prestudy-interface}). We randomized the presentation order of these claims to prevent order effects. Based on the plausibility of each claim, participants selected one of the three options: \textit{Reasonable}, \textit{Unreasonable}, and \textit{Not Sure}. As each participant rated more than a dozen causal claims, we used the three options rather than a Likert scale with five options or more to keep the study short. During the study, participants also answered two attention check questions asking them to directly select one of the three options.

\subsection{Results}

For each causal claim, we computed the probabilities that participants selected \textit{Reasonable}, \textit{Unreasonable}, and \textit{Not Sure}. We then calculated the entropy for each claim. A low entropy implies that participants mostly voted for the same option, whereas a high entropy means that participants' votes tended to distribute across the three options. Within each bucket of the 30 reasonable claims, 30 unreasonable claims, and 30 hard-to-tell claims, we ranked the claims by entropy. 

For the 30 reasonable claims, we ranked them in increasing order of entropy. The top claims had a low entropy because participants mostly voted for \textit{Reasonable}. For the 30 unreasonable claims, we again ranked them in increasing order of entropy. Participants mostly selected \textit{Unreasonable} for the top claims. For the 30 hard-to-tell claims, we expected that the claims where the plausibility was the most difficult to judge had a high entropy score. This is because participants likely struggled to choose among the options. We sorted these claims in decreasing order of entropy.

Figure~\ref{topClaims} shows the top three reasonable, unreasonable, and hard-to-tell claims based on the above ranking. We provide a ranked list of all the 90 claims as a supplementary material. Study 1 confirmed the validity of our results: In study 1, when participants rated the claims on a 7-point Likert scale, they tended to agree with the top reasonable claims, disagree with the top unreasonable claims, and be neutral about the top hard-to-tell claims.

\begin{figure*}[t!]
	\centering
	\includegraphics[width=\linewidth]{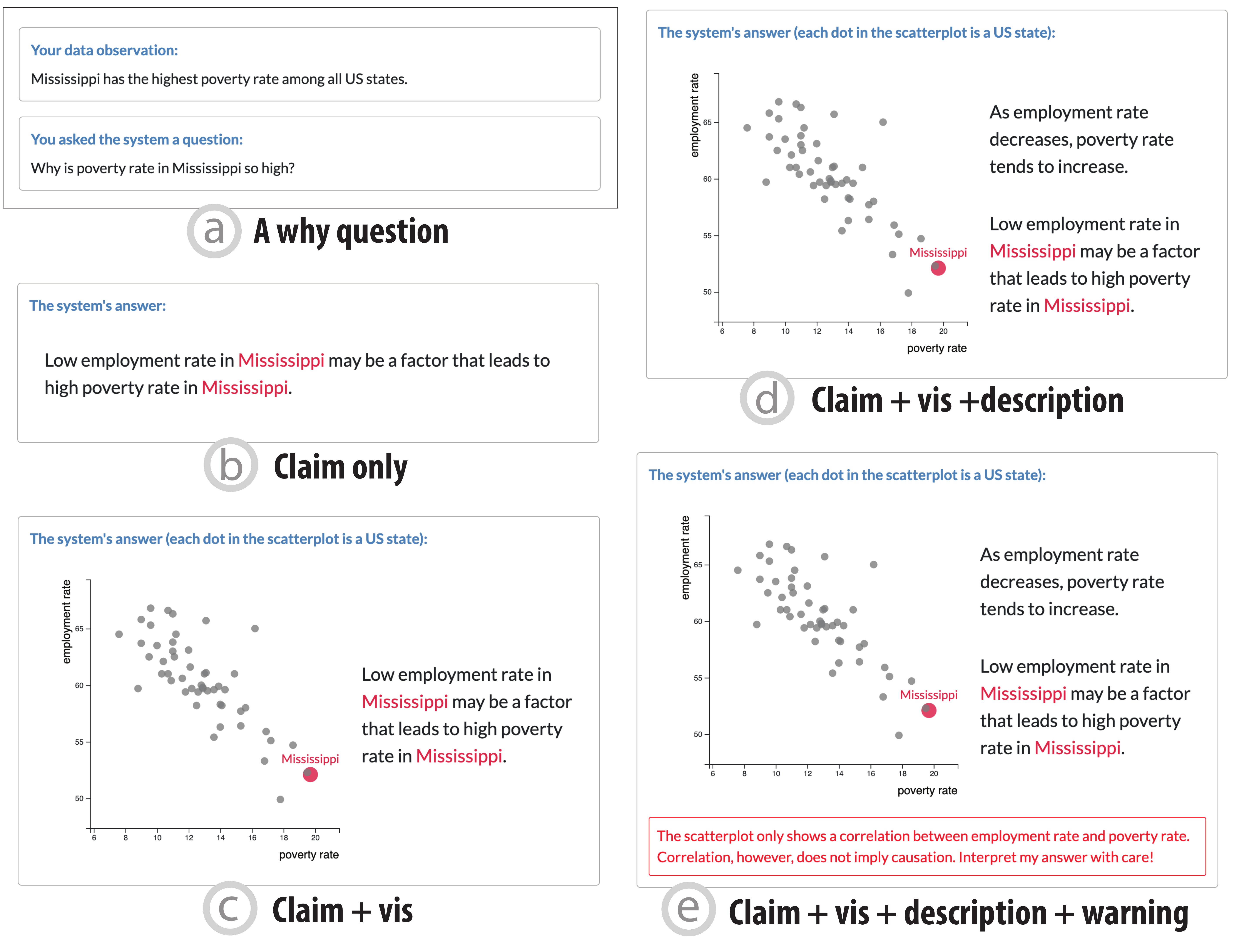}
	\caption{The four experimental conditions. A user asks about the high poverty rate in Mississippi (a). The system answers only a causal claim (b), shows a scatterplot next to the claim (c), adds a description about the correlation (d), and warns about the system's flaws besides showing the previous information (e).}
	\Description{Screenshots of the four experimental conditions. The claim-only condition shows only the causal claim ``Low employment rate in Mississippi may be a factor that leads to high poverty rate in Mississippi.'' The claim + vis condition shows a scatterplot with poverty rate on the X axis and employment rate on the Y axis. In the plot, poverty rate and employment rate are negatively correlated. The claim + vis + description condition shows the description ``As employment rate decreases, poverty rate tends to increase.'' next to the causal claim and the scatterplot. The claim + vis + description + warning condition shows a message ``The scatterplot only shows a correlation between employment rate and poverty rate. Correlation, however, does not imply causation. Interpret my answer with care!''}
	\label{conditions}
\end{figure*}

\section{Study 1: Providing Answers with Different Plausibility}

With the ranked lists of reasonable, unreasonable, and hard-to-tell claims, we designed a between-subject experiment during which participants reviewed a series of answers to why questions.

When designing the presentation of answers, we considered its complexity to typical end users. One way to answer why questions (e.g., why Mississippi has a high poverty rate) is causal graph~\cite{causalInference, cGraph}, a technique often employed in statistics literature for visualizing complex causal relationships. However, causal graphs may require more advanced statistics training to understand.

We also considered showing multiple factors in the answer but were concerned about introducing experimental confounds. For example, the number, the perceived plausibility, and the underlying causal relationship of the factors could potentially alter the perception of system performance. Yet, using real-world data implied that these variables could be difficult to control for. 

We therefore adopted a simplified design where the system responded to a why question by stating a factor that could answer the question. In each task, participants saw a why question (e.g., why Mississippi has a high poverty rate) and the system's answer (e.g., low employment rate in Mississippi may be a factor that leads to the high poverty rate). Across conditions, the answers had different designs (Fig.~\ref{conditions}b-d). We provide screenshots of the experiment interface as a supplementary material.

\begin{figure*}[t!]
	\centering
	\includegraphics[width=\linewidth]{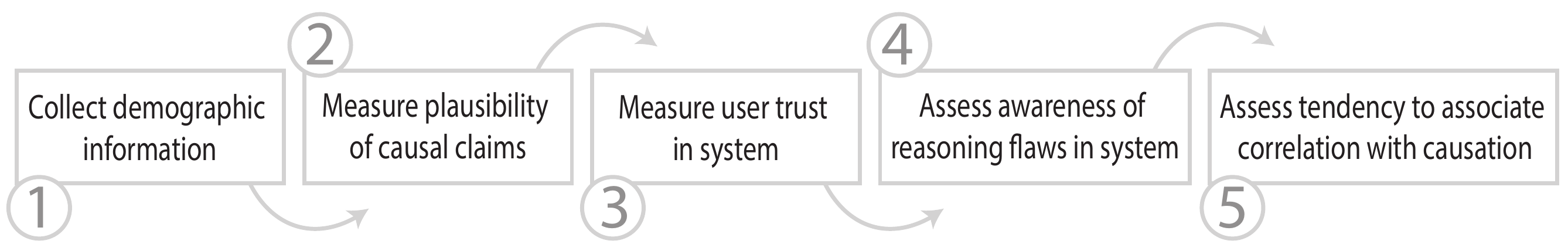}
	\caption{The five stages in study 1 and study 2.}
	\Description{A flow chart showing the five stages in the experiment. In the figure, the five stages are ``collect demographic information,'' ``measure plausibility of causal claims,'' ``measure user trust in system,'' ``assess awareness of reasoning flaws in system,'' and ``assess tendency to associate correlation with causation.''}
	\label{flow}
\end{figure*}

\subsection{Methods}

\subsubsection{Conditions}

Building on prevailing system designs and the research literature, we focused on two types of correlational evidence (scatterplot and textual description about correlation) to investigate whether they created an illusion of causality. We further studied the effectiveness of warning in reducing the illusion. Here, we describe these three types of information:

\vspace{1mm} \noindent \textbf{Scatterplot}. Scatterplots are common for showing the relationship between two numerical variables~\cite{scatterDesign}. They have also been applied in question-answering functionality in commercial systems for showing the relationship between cause and effect (Fig.~\ref{explain}).

\vspace{1mm} \noindent \textbf{Textual description about correlation}. While the causal claim (e.g., low employment rate in Mississippi may be a factor that leads to the high poverty rate) describes a single state in the US, a description about correlation (e.g., as employment rate decreases, poverty rate tends to increase) depicts the overall trends for all the states. To facilitate interpretation, visualization systems often provide such descriptions next to a chart~\cite{characterizing, voder}.

\vspace{1mm} \noindent \textbf{Warning message}. Although scatterplots and the textual descriptions only reveal correlation, they might induce an illusion of causality~\cite{correlation}. A mitigation strategy is to use a message to warn users that correlation is not causation. While such warnings are less common in visualization systems, they are commonly used in other systems (e.g., web browser) to prompt safety-related behaviors (e.g., not to click on phishing websites)~\cite{phishing}. It would be interesting to learn about if a simple statement is enough to raise awareness of the system's potential flaws and reduce users' tendency to confuse correlation and causation.

Based on these three information types, we designed four answer interfaces by adding the information types one by one. Participants were randomly assigned to a condition where the answers adopted one of the four designs:

\vspace{1mm} \noindent \textbf{Claim only} (Fig.~\ref{conditions}b). The system only shows a claim about cause and effect as an answer to a why question.

\vspace{1mm} \noindent \textbf{Claim + vis} (Fig.~\ref{conditions}c). Beside the causal claim, the system visualizes the cause and effect using a scatterplot. Prior studies showed that the aspect ratio of point clouds in a scatterplot affects correlation estimation~\cite{cleveland, scatterplotOp}. To support a consistent correlation estimation, we controlled the aspect ratio. For each axis, we set the lowest value to be (min value of the data $-$ 0.15 $\times$ range of the data) and the highest value to be (max $+$ 0.15 $\times$ range). 

\vspace{1mm} \noindent \textbf{Claim + vis + description} (Fig.~\ref{conditions}d). The system additionally states the correlation between the cause and effect variables with a textual description.

\vspace{1mm} \noindent \textbf{Claim + vis + description + warning} (Fig.~\ref{conditions}e). To encourage users to evaluate the answers carefully, the system warns that the scatterplot only shows correlation, and that correlation is not causation.

\subsubsection{Participants}

A power analysis indicated that for a significance level of 0.05 and a power of 0.8, detecting a medium effect size of $f = 0.25$ using one-way ANOVA required 180 participants (45 participants per condition). As we planned to conduct non-parametric tests (see Sec.~\ref{quantitative}), we targeted a slightly larger sample size (200 participants in total or 50 participants per condition) following guidelines on sample size determination for non-parametric tests~\cite{nonparametric}.

During participant recruitment, we limited the study to workers in the United States, had an acceptance rate of 95\% or above, and did not participate in the pre-study. The study took approximately 10-20 minutes, and we compensated participants \$2.90. At the end, we recruited 200 unique workers on MTurk.  

The survey had two interpretation checks for assessing scatterplot comprehension and three open-ended questions (details in the Procedure section). We excluded participants who did not pass any of the interpretation checks or provided gibberish answers for any of the open-ended questions (but compensated them for participation). Overall, the data quality was poor. For example, many participants provided canned responses for some open-ended questions. We omitted 123 participants and continued recruiting until reaching the target sample size.

Participants aged 20-69 (\textit{M}=35.4, \textit{SD}=10.1). 131 were male, 68 were female, and 1 preferred not to say. They reported different educational attainments: high school (29 participants), professional school (22), college (109), graduate school (35), PhD (1), and postdoctoral (4). Concerning data analysis expertise, 44 had none, 64 were beginners, 71 were intermediate, and 21 were advanced. For experience with visualization platforms (e.g., Tableau), 82 had none, 60 were beginners, 36 were intermediate, and 22 were advanced. When asked about the frequency of using question-answering systems, 132 reported never, 30 reported rarely, 23 reported weekly, and 15 reported daily.

\begin{figure*}[t!]
	\centering
	\includegraphics[width=\linewidth]{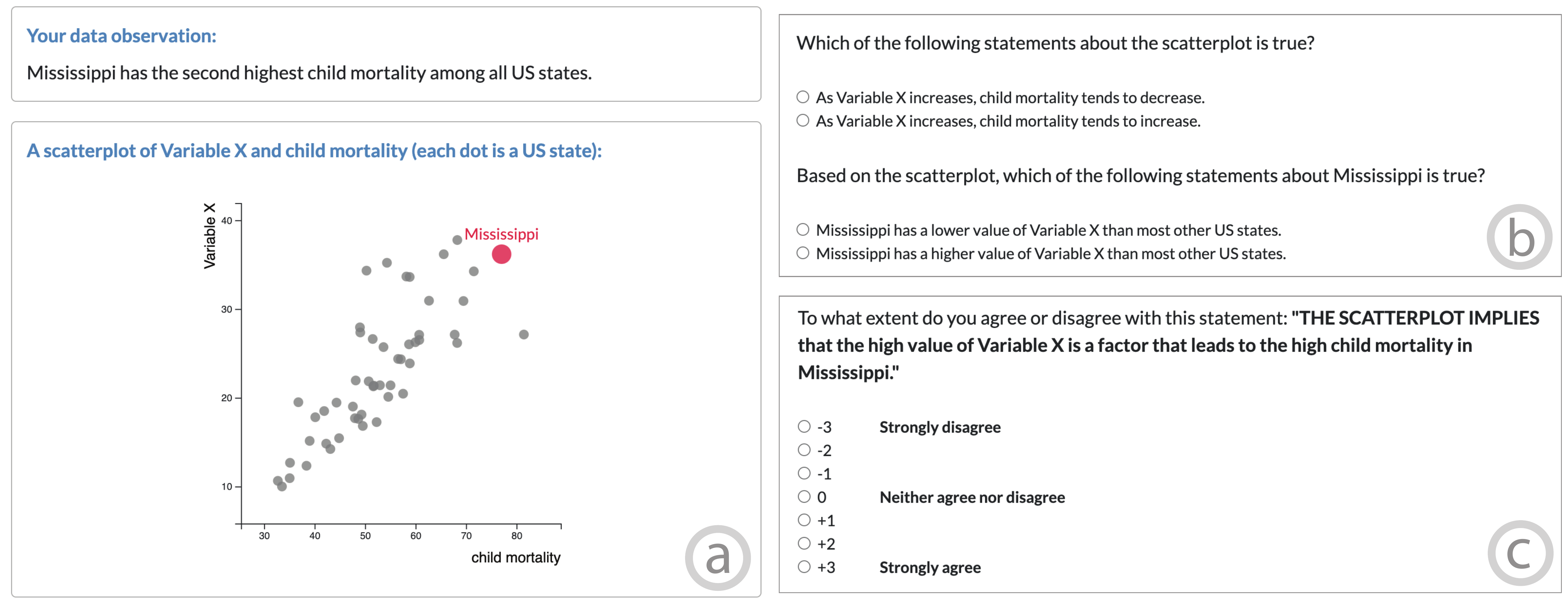}
	\caption{Measuring tendency to associate correlation with causation. Participants saw a data observation and a scatterplot (a), answered interpretation check questions (b), and rated their agreement on a statement suggesting that the scatterplot implied causation (c).}
	\Description{Screenshots of the survey section for assessing awareness of ``correlation is not causation.'' First, participants saw a statement ``Mississippi has the second highest child mortality among all US states.'' Below the statement is a scatterplot with child mortality on the X axis and Variable X on the Y axis. The plot shows a positive correlation between the two variables. From the plot, Mississippi has a high value in both child mortality and Variable X. Based on the scatterplots, participants were asked two interpretation check questions. The first question says, ``Which of the following statements about the scatterplot is true?'' There are two options: ``As Variable X increases, child mortality tends to decrease,'' and ``As Variable X increases, child mortality tends to increase.'' The second question says, ``Based on the scatterplot, which of the following statements about Mississippi is true?'' There are two options: ``Mississippi has a lower value of Variable X than most other US states,'' and ``Mississippi has a higher value of Variable X than most other US states.'' After the interpretation checks, participants were asked, ``To which extent do you agree or disagree with this statement: `The scatterplot implies that the high value of Variable X is a factor that leads to the high child mortality in Mississippi.''' Below this question, they could answer by selecting a value from -3 (meaning strongly disagree) to +3 (meaning strongly agree).}
	\label{awareness}
\end{figure*}

\subsubsection{Procedure}

We first randomly assigned participants to one of the four conditions. For all conditions, the study consisted of five main stages (Fig.~\ref{flow}).

In stage 1 (Fig.~\ref{flow} \circled{1}), participants filled out a demographic survey. After filling out the survey, they completed a practice task to get acquainted with the study interface.

In stage 2 (Fig.~\ref{flow} \circled{2}), participants reviewed a series of nine answers to why questions. In each task, they examined a data observation (e.g., Mississippi has the highest poverty rate among all US states) (Fig.~\ref{conditions}a), a why question (e.g., why is poverty rate in Mississippi so high?) (Fig.~\ref{conditions}a), and the system’s answer to the question (Fig.~\ref{conditions}b-d). Depending on the condition, participants saw a different visual design for the answers. Based on the system’s answer, participants rated their agreement with a causal claim (e.g., low employment rate in Mississippi is a factor that leads to high poverty rate in Mississippi) on a 7-point Likert scale.

We constructed the nine answers using the top nine causal claims obtained from the pre-study (Fig.~\ref{topClaims}). Hence, three answers were reasonable, three were unreasonable, and three had plausibility that was difficult to judge. This intended to mirror real-world systems that tend to be unreliable in answering why questions. The order of the answers was randomized to prevent order effects.

After participants reviewed the nine answers, we measured user trust in the system in stage 3 (Fig.~\ref{flow} \circled{3}). Participants rated their trust in the system on a 7-point scale from -3 (I don't trust it at all) to +3 (I fully trust it). They further shared their reasons for trusting or not trusting the system.

Next, we assessed their awareness of the reasoning flaws in the system in stage 4 (Fig.~\ref{flow} \circled{4}). Participants reported whether they observed any reasoning flaws in the system. If the answer was ``yes,'' we asked them to specify the reasoning flaw(s) they found.

In the final stage (Fig.~\ref{flow} \circled{5}), we assessed their tendency to associate correlation with causation. Participants in stage 5 saw a description of a data observation (Mississippi has the second highest child mortality among all US states) and a scatterplot showing a strong correlation between child mortality and an unknown variable X (Fig.~\ref{awareness}a). To assess participants' understanding of scatterplots, we first asked participants to answer two interpretation check questions (Fig.~\ref{awareness}b). Participants who failed to pass any of the questions were excluded from the data analysis.

Whereas participants rated their agreement with a causal relationship in stage 1, participants rated their agreement with a sentence stating that a scatterplot with a high correlation implied a causal relationship in stage 5. Participants saw a statement: ``\textit{The scatterplot implies} that the high value of Variable X is a factor that leads to the high child mortality in Mississippi'' (Fig.~\ref{awareness}c). They rated the statement on a 7-point Likert scale and explained why they agreed or disagreed. 

\subsubsection{Quantitative Measures}
\label{quantitative}

We derived six measures from participants' response.

\vspace{1mm} \noindent \textbf{Agreement (reasonable)}. For each participant, we computed the average agreement rating for the three reasonable answers.

\vspace{1mm} \noindent \textbf{Agreement (unreasonable)}. It is the average rating for the three unreasonable answers.

\vspace{1mm} \noindent \textbf{Agreement (hard to tell)}. It is the average rating for the three answers that were hard to tell if they were reasonable.

\vspace{1mm} \noindent \textbf{Trust}. Some researchers have developed questionnaires to assess user trust in recommender systems~\cite{recommender} and machine learning systems~\cite{ml}. Since these questionnaires may not be applicable to question-answering systems, we tailored a question to assess trust in question-answering systems. In the post-study survey, we asked, ``Overall, how much do you trust or not trust the question-answering system?'' and participants rated on a scale from -3 to +3. 

\vspace{1mm} \noindent \textbf{Awareness of system's flaws}. We computed the number of participants who selected ``yes'' for the question, ``Did you observe any flaw(s) in the reasoning of the question-answering system?'' Unlike the other measures that are scales between -3 and +3, this measure is a count between zero and 50. Whereas trust and agreement with answers are more subjective, observations about reasoning flaws in the system are more clear-cut, making a yes/no question more suitable.

\vspace{1mm} \noindent \textbf{Awareness of ``correlation is not causation''.} In the last part, participants rated a statement: ``\textit{The scatterplot implies} that the high value of Variable X is a factor that leads to the high child mortality in Mississippi.'' (Fig.~\ref{awareness}c) If participants were cautious about drawing causal conclusions from correlation, they should be inclined to disagree with the statement.

During a pilot study, we observed that when the variable name was shown, participants tended to use their common sense to decide if they agreed with the statement. Yet, we wanted to assess tendency to confuse correlation and causation instead of ability to apply common sense. To reduce the impact of common sense in answering the question, we hid the variable name of X.

Likert-scale data are not continuous and violate the ANOVA assumptions. To study the main effect of answer design, we used a Kruskal-Wallis test, which is a non-parametric equivalence of one-way ANOVA, for the five measures using a 7-point scale (i.e., all measures except awareness of system's flaws). When there is a significant main effect, we conducted post-hoc Wilcoxon rank sum tests with a Holm-Bonferroni correction for pairwise comparisons. 

For the awareness of system's flaws, we used a Fisher's exact test to assess if the number of participants who found reasoning flaws in the system was significantly different across conditions. 

\begin{figure*}[t!]
	\centering
	\includegraphics[width=\linewidth]{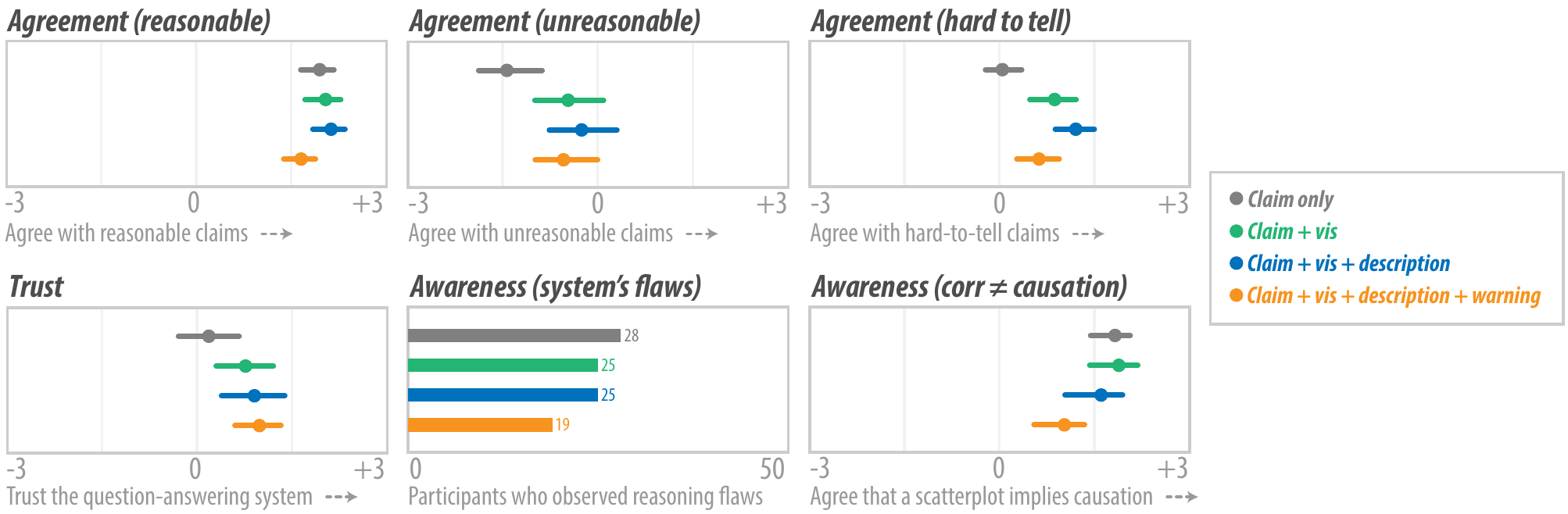}
	\caption{Quantitative results from study 1. All error bars show 95\% bootstrapped confidence intervals.}
	\Description{Six charts showing the quantitative results for the six quantitative measures in study 1. Each chart shows the measures for the four conditions. For example, in the chart for the measure ``awareness of system’s flaws,'' there are 28, 25, 25, and 19 participants who observed reasoning flaws in the system for the claim-only, claim + vis, claim + vis + description, and claim + vis + description + warning conditions respectively. Other numbers can be found in the text.}
	\label{study1}
\end{figure*}

\subsubsection{Qualitative Response}

There were three open-ended questions in the survey, one for explaining trust or distrust, one for specifying reasoning flaws in the system, and one for explaining why agreed or disagreed that the scatterplot implied causation.

For each question, an author open-coded the responses to identify the emergent categories and develop a codebook. We observed that a response could include multiple categories. Hence, we treated each category as binary: For each response, we labelled whether each category was present or absent. Two coders independently coded all responses. We then discussed inconsistencies, refined code definitions, and independently re-coded the responses based on the new definitions. We iteratively coded the responses until we reached a Cohen's $\kappa$ above 0.7 for all the categories.

For each category, we conducted a Fisher's exact test to determine whether its presence was significantly different across conditions.

\subsubsection{Hypotheses}

We developed hypotheses based on research in visualization's persuasive power, trust in automated systems, and warning science.

Pandey et al.~\cite{persuasive} found that when participants did not have a strong attitude towards a topic, visualizations had a strong power to change their attitudes. They also commented on the difficulty to change attitudes for topics of which participants already had a strong prior opinion~\cite{persuasive}. We expected that showing a scatterplot would increase the plausibility of hard-to-tell claims because participants likely did not have a strong attitude towards them. We also expected that the scatterplot would not affect the plausibility of reasonable and unreasonable claims.

\vspace{1mm} \noindent \textbf{H1.1}: Participants' agreement with the reasonable claims does not differ across conditions.

\vspace{1mm} \noindent \textbf{H1.2}: Participants' agreement with the unreasonable claims does not differ across conditions.

\vspace{1mm} \noindent \textbf{H1.3}: Participants in the three conditions that show a scatterplot in the answers (i.e., claim + vis, claim + vis + description, and claim + vis + description + warning) agree with the hard-to-tell claims more than participants in the claim-only condition.

Transparency in automated systems can inspire user trust~\cite{transparency}. For example, when a recommender system provides reasons behind its recommendations, users tend to trust the system more~\cite{cfs}. Showing the data can increase the transparency in the question-answering system. We posited that users would trust the system more when it showed the scatterplot.

\vspace{1mm} \noindent \textbf{H1.4}: Participants in the three conditions that show a scatterplot in the answers trust the question-answering system more than participants in the claim-only condition.

Some researchers in warning science have compared the effectiveness of passive and active warnings~\cite{phishing}. Whereas active warning forces users to notice it by blocking user tasks, passive warning (e.g., a simple warning message) is less interrupting~\cite{phishing}. In data analysis, passive warning is more suitable because a small latency in interaction can hamper analysis quality~\cite{latency}. However, Egelman et al.~\cite{phishing} showed that passive warnings were often ineffective because users might ignore them. The ineffectiveness might extend to question-answering systems. Hence, we posited that the warning message would not increase participants' awareness of the system's flaws nor decrease their tendency to associate correlation with causation.

\vspace{1mm} \noindent \textbf{H1.5}: Participants' awareness of the system's flaws does not differ across conditions.

\vspace{1mm} \noindent \textbf{H1.6}: Participants' awareness of ``correlation is not causation'' does not differ across conditions.

\subsection{Results}

Figure~\ref{study1} summarizes the results for the quantitative measures. We observed that the scatterplot increased the plausibility of unreasonable and hard-to-tell claims but not reasonable claims. The warning message appeared to decrease the plausibility of reasonable claims but not unreasonable and hard-to-tell claims. Trust and awareness of flaws did not seem to differ across conditions. However, the warning message seemed to increase the awareness of ``correlation is not causation.''

Also, for the claim-only condition, participants tended to give a neutral rating for the hard-to-tell claims (\textit{M}=0.04), a positive rating for the reasonable claims (\textit{M}=1.95), and a negative rating for the unreasonable claims (\textit{M}=-1.44). This confirmed the validity of the pre-study results. 

In the following, we provide the detailed analysis.

\subsubsection{Agreement (reasonable)}

On a scale from -3 (strongly disagree) to +3 (strongly agree), participants in the claim + vis + description condition rated the reasonable claims the highest (\textit{M}=2.13, \textit{SD}=0.92), followed by those in the claim + vis condition (\textit{M}=2.04, \textit{SD}=0.98), the claim-only condition (\textit{M}=1.95, \textit{SD}=0.96), and the claim + vis + description + warning condition (\textit{M}=1.65, \textit{SD}=0.89). A Kruskal-Wallis test indicated a significant main effect of answer design on the rating ($\chi^2(3)$=10.8, \textit{p}=.013). We conducted six post-hoc pairwise comparisons using Wilcoxon rank sum tests with a Holm-Bonferroni correction. Results showed that only the difference between claim + vis + description + warning and claim + vis + description (\textit{p}=.017) as well as that between claim + vis + description + warning and claim + vis (\textit{p}=.044) were significant. The results did not support \textbf{H1.1}.

\subsubsection{Agreement (unreasonable)}

Participants in the claim + vis + description condition rated the unreasonable claims the highest (\textit{M}=-0.26, \textit{SD}=1.92), followed by those in the claim + vis condition (\textit{M}=-0.47, \textit{SD}=1.96), the claim + vis + description + warning condition (\textit{M}=-0.55, \textit{SD}=1.82), and finally the claim-only condition (\textit{M}=-1.44, \textit{SD}=1.74). A Kruskal-Wallis test indicated a significant main effect of answer design on the rating ($\chi^2(3)$=12.2, \textit{p}=.007), with post-hoc pairwise comparisons showing that all the three conditions with scatterplots in the answers had a significantly higher average rating than the claim-only condition. The results did not support \textbf{H1.2}.

\subsubsection{Agreement (hard to tell)}

Participants in the claim + vis + description condition rated the hard-to-tell claims the highest (\textit{M}=1.2, \textit{SD}=1.12), followed by those in the claim + vis condition (\textit{M}=0.87, \textit{SD}=1.34), the claim + vis + description + warning condition (\textit{M}=0.62, \textit{SD}=1.25), and the claim-only condition (\textit{M}=0.04, \textit{SD}=1.05). There is a significant main effect of answer design on the rating ($\chi^2(3)$=24.4, \textit{p}<.001). Pairwise comparisons showed that all the three conditions with scatterplots in the answers had a significantly higher average rating than the claim-only condition. Other pairs were not significantly different. The findings supported \textbf{H1.3}.

\subsubsection{Trust}

On average, the trust ratings across conditions were positive, indicating a tendency to trust the system. Claim + vis + description + warning has the highest rating (\textit{M}=0.98, \textit{SD}=1.31), followed by claim + vis + description (\textit{M}=0.9, \textit{SD}=1.76), claim + vis (\textit{M}=0.76, \textit{SD}=1.67), and claim-only (\textit{M}=0.18, \textit{SD}=1.70). However, we did not observe a significant main effect of answer design on trust ($\chi^2(3)$=7.34, \textit{p}=.062). The results did not support \textbf{H1.4}.

\subsubsection{Why trust or not trust?}

We coded participants' reasons for trusting or not trusting the question-answering system. Seven categories of responses emerged from the analysis. We report the core results here and provide the detailed breakdown of the categories across conditions in the supplementary materials. 

For each response, we labelled each category as present or absent. We labelled all categories as absent for responses that were too broad or vague (e.g., \textit{``it is nice''}). 

The top three reasons for distrusting the system were some answers did not make sense (40.5\% of 200), the system confused correlation and causation (9\%), and it did not provide enough support for its causal claims (7.5\%). A participant felt that some claims lacked support and wrote, \textit{``Some of the answers could be factual but it was hard to determine without further data.''}

The top three reasons for trusting the system were that some answers made sense (26\%), the system showed the data (8.5\%), and the system provided some support for its causal claims (6.5\%)

We did not observe a significant difference in the presence of any of the categories across conditions using Fisher's exact tests (details in supplementary materials).

\subsubsection{Awareness of system's flaws}

Using a Fisher's exact test, we did not find a significant difference in the number of people who found reasoning flaws (these participants selected ``yes'' for the question asking whether they observed reasoning flaws) across conditions (\textit{p}=.33). We could not reject \textbf{H1.5}.

\subsubsection{What are the reasoning flaws?}

The qualitative coding resulted in four categories. Among the 97 participants who observed reasoning flaws in the system, the majority of participants stated providing nonsensical answers as a reasoning flaw (70.1\% of 97). Other observed reasoning flaws were confusing correlation and causation (15.5\%), not having enough support for the claims (8.25\%), and considering only one factor (3.09\%). Fisher's exact tests did not indicate significant differences in the presence of any of the four categories across conditions.

\subsubsection{Awareness of ``correlation is not causation''}

We asked participants to rate a sentence stating that a scatterplot implied causation. On a scale from -3 (strongly disagree) to +3 (strongly agree), claim + vis + description + warning had the lowest rating (\textit{M}=1.02, \textit{SD}=1.45), followed by claim + vis + description (\textit{M}=1.6, \textit{SD}=1.59), claim-only (\textit{M}=1.82, \textit{SD}=1.10), and claim + vis (\textit{M}=1.88, \textit{SD}=1.33). All conditions got a positive average rating, indicating a tendency to associate correlation with causation. We found a significant main effect of answer design on the rating ($\chi^2(3)$=15.2, \textit{p}=.002). Post-hoc pairwise comparisons showed that claim + vis + description + warning had a significantly lower average rating than all the other three conditions, indicating that the warning appeared to reduce the tendency to associate correlation with causation. The results did not support \textbf{H1.6}.

\subsubsection{Why agree or disagree with the statement?}

The qualitative coding yielded four categories. We again observed that some responses were overly broad (e.g., \textit{``because the graph shows it''}) and coded all categories as absent for such responses. 

Among the more specific responses, the majority of participants agreed that the scatterplot implied a causal relationship because the scatterplot showed a correlation (46\% of 200). An example response is \textit{``If Variable X did not rise then child mortality would not rise.''}

Participants disagreed with the statement because correlation is not causation (8.5\%), variable X was unknown and they could not judge (8.5\%), and the scatterplot had outliers (4\%). A participant who observed outliers said, \textit{``I only slightly agree because other states show otherwise. Texas, for instance, has a much lower Child Mortality rate but Variable X is almost the same.''}

We did not find significant differences in the presence of any of the categories across conditions.

\begin{figure*}[t!]
	\centering
	\includegraphics[width=0.75\linewidth]{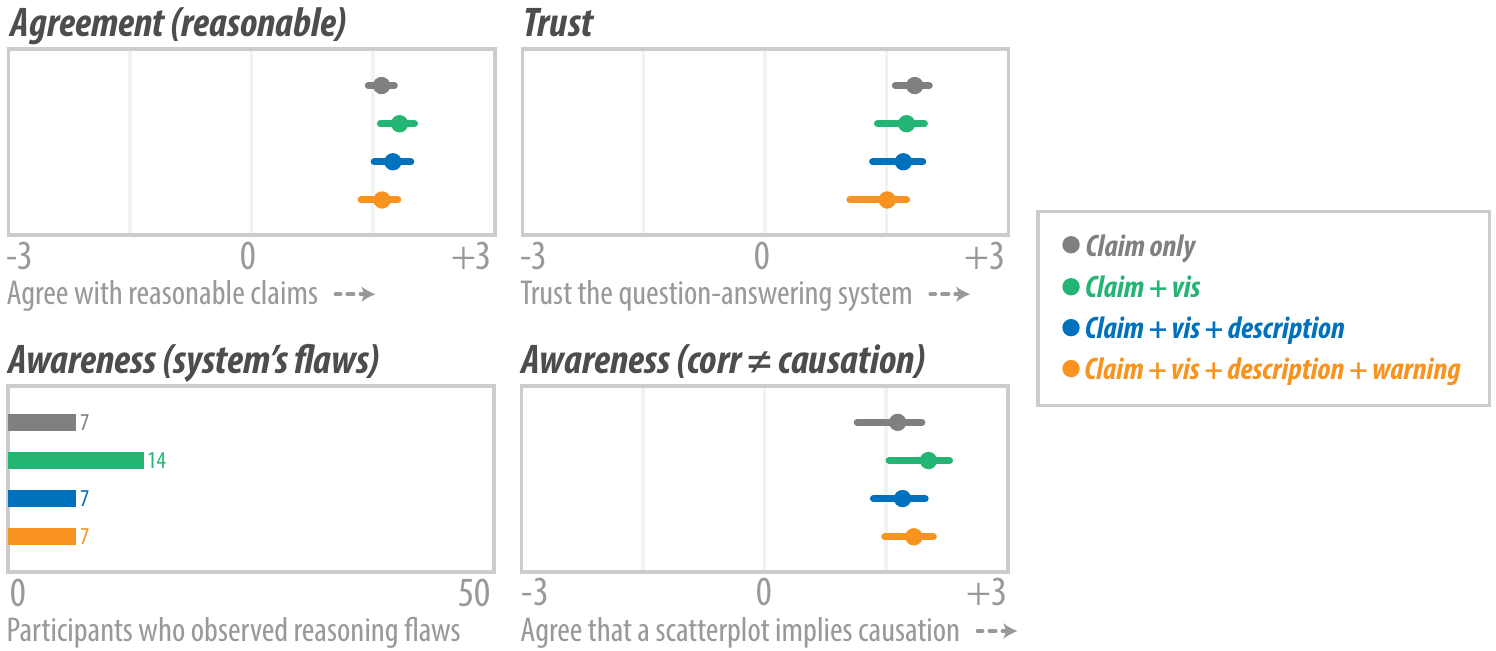}
	\caption{Quantitative results from study 2. All error bars show 95\% bootstrapped confidence intervals.}
	\Description{Four charts showing the quantitative results for the four quantitative measures in study 2. Each chart shows the measures for the four conditions. For example, in the chart for the measure ``awareness of system’s flaws,'' there are 7, 14, 7, and 7 participants who observed reasoning flaws in the system for the claim-only, claim + vis, claim + vis + description, and claim + vis + description + warning conditions respectively. The three other charts are agreement with reasonable claims, trust, and awareness of ``correlation is not causation.'' These three charts do not show statistically significant differences across the four conditions.}
	\label{study2}
\end{figure*}

\section{Study 2: Providing Only Reasonable Answers}

Several findings from study 1 deviated from our expectations: The simple warning appeared to decrease the plausibility of reasonable claims and increase the awareness of ``correlation is not causation''; we did not have enough evidence that user trust was improved by showing the data. A potential explanation lied in the unreliable performance of the system---it made the warning more noticeable and reduced the effectiveness of showing the data in improving user trust (when the system performed poorly, it was untrustworthy no matter whether it showed the data). To investigate whether the observations in study 1 held for a system that had a higher perceived performance, we conducted study 2.

\subsection{Methods}

Study 2 was the same as the study 1 except that participants reviewed nine reasonable answers to why questions (as opposed to reviewing answers with different levels of plausibility in study 1). We constructed the answers using the top nine claims in the ranked list of 30 reasonable claims obtained from the pre-study.

We similarly recruited 50 participants per condition (200 unique workers in total). Workers who participated in the pre-study and study 1 were excluded from study 2. Participants aged 18-70 (\textit{M}=36.1, \textit{SD}=11.0). 121 were male, 77 were female, and 2 preferred not to say. The reported educational attainments were high school (28 participants), professional school (10), college (116), graduate school (37), PhD (8), and postdoctoral (1). Concerning data analysis expertise, 44 had none, 79 were beginners, 54 were intermediate, and 23 were advanced. For experience with visualization platforms (e.g., Tableau), 84 had none, 47 were beginners, 45 were intermediate, and 24 were advanced. When asked about the frequency of using question-answering systems, 123 reported never, 35 reported rarely, 31 reported weekly, and 11 reported daily.

As the system only presented reasonable answers, study 2 only had four measures: agreement (reasonable), trust, awareness of system's flaws, and awareness of ``correlation is not causation.''

In study 1, participants heeded the warning, causing them to agree less with reasonable claims and be less likely to associate correlation with causation. We expected that both effects would disappear when the system was more trustworthy. Furthermore, in study 1, showing the data using a scatterplot did not seem to improve user trust in the system. We posited that when the system provided only reasonable answers, showing the data would improve user trust. We considered the same set of hypotheses as in study 1:

\vspace{1mm} \noindent \textbf{H2.1}: Participants' agreement with the reasonable claims does not differ across conditions.

\vspace{1mm} \noindent \textbf{H2.2}: Participants in the three conditions that show a scatterplot in the answers trust the question-answering system more than participants in the claim-only condition.

\vspace{1mm} \noindent \textbf{H2.3}: Participants' awareness of the system's flaws does not differ across conditions.

\vspace{1mm} \noindent \textbf{H2.4}: Participants' awareness of ``correlation is not causation'' does not differ across conditions.

\subsection{Results}

Figure~\ref{study2} shows the quantitative results. Kruskal-Wallis tests for agreement (reasonable), trust, and awareness of ``correlation is not causation'' as well as a Fisher's exact test for awareness of system's flaws indicated no significant differences across conditions (details in the supplementary materials). Hence, the results failed to support \textbf{H2.2}. However, we could not reject \textbf{H2.1}, \textbf{H2.3}, and \textbf{H2.4}.

We also observed that participants in study 2 appeared to trust the system more than those in study 1. The mean trust rating in study 2 was 1.70 (\textit{SD}=1.04) while that in study 1 was 0.71 (\textit{SD}=1.64). Participants in study 2 also found fewer reasoning flaws in the system. The total number of participants who found reasoning flaws in study 2 was 35 (compared with 97 in study 1). We summarize the qualitative results as follows.

\subsubsection{Why trust or not trust?}

Participants provided diverse reasons for trusting or not trusting the system. Seven categories of reasons emerged from the qualitative coding.

The top three reasons for trusting the system were the answers made sense (38.5\% of 200), the system provided enough support for its causal claims (19.5\%), and it showed the data (8\%). The top three reasons for distrusting the system were the system did not provide enough support for its claims (8\%), it considered only one factor (7\%), and it confused correlation and causation (2.5\%). 

Fisher's exact tests indicated that the number of participants stating ``the answers made sense'' as a reason was significantly different across conditions (\textit{p}=.014). We conducted six post-hoc pairwise comparisons using Fisher’s exact tests with a Holm-Bonferroni correction. We only observed that more participants in the claim-only condition stated ``the answers made sense'' than in the claim + vis + description condition (\textit{p}=.025). A potential explanation was that providing a claim only led participants to comment mostly on the plausibility of the claim. However, providing other information (e.g., a scatterplot) alongside a claim enabled them to comment on other aspects and less on the plausibility.

In study 2, 56\% of the responses contained reasons for trusting the system while 16\% contained reasons for not trusting it. The data stood in contrast to those in study 1. In study 1, 39.5\% of the responses had reasons for trust while 49.5\% had reasons for distrust. This echoed the finding that participants trusted the system more in study 2.

\subsubsection{What are the reasoning flaws?}

Among the 35 participants who answered ``yes'' for the question asking whether they observed reasoning flaws in the system, we found three categories of responses after omitting those who provided vague answers: The system considered only one factor (28.6\% of 35); it confused correlation and causation (20\%); it did not provide enough support for the claims (17.1\%). Using Fisher's exact tests, we did not observe significant differences in the presence of the categories across conditions.

\subsubsection{Why agree or disagree with the statement?}

The qualitative analysis resulted in five categories of responses. Congruent with study 1's results, most participants agreed that the scatterplot implied a causal relationship because the scatterplot showed a correlation (43.5\% of 200).

Participants who disagreed with the statement commented that correlation is not causation (7.5\%), the scatterplot had outliers (7\%), variable X was unknown and they could not judge (5.5\%), and the dots in the scatterplot looked disperse (1.5\%).

Fisher's exact tests did not show a significant difference across conditions for any of the categories.

\section{Discussion}

Before discussing the implications of our findings, we summarize the results from the two studies and provide potential explanations for the less intuitive observations.

In study 1, participants reviewed answers of different plausibility. We did not observe effects of the textual description about correlation on the perceived plausibility of causal claims, user trust in the system, the awareness of the system's flaws, and the awareness of ``correlation is not causation.'' However, showing a scatterplot caused participants to disagree less with unreasonable claims and agree more with hard-to-tell claims. In contrast, a simple warning message seemed to cause participants to agree less with reasonable claims. The warning also reduced participants' tendency to associate correlation with causation.

Nevertheless, when participants examined only reasonable answers in study 2, the impact of the simple warning message on reducing the plausibility of reasonable claims and on raising the awareness of ``correlation is not causation'' seemed to disappear. Research in warning science found that arousal strength (i.e., the perceived importance or relevance of a warning) affects the effectiveness of a warning message in motivating safety-related behaviors~\cite{arousal}. Participants in study 2 tended to trust the system more than those in study 1. This likely led participants in study 2 to perceive the warning about the system's reasoning flaws to be less relevant. The warning in study 2 became less effective possibly because participants tended to ignore the warning.

In both studies, we did not observe significant differences in user trust and the awareness of the system's flaws across conditions. The qualitative results provided an explanation. In study 1, when asked about why they did not trust the system or what were the reasoning flaws in the system, most participants simply stated that the answers did not make sense. In study 2, when asked about why they trusted the system, the majority commented that the answers made sense to them. The results appeared to indicate that system performance in answering why questions had a dominating effect on user trust and the awareness of reasoning flaws in the system. In other words, when users can assess system performance, showing other information (e.g., a scatterplot or a warning) may play a small role in shaping user trust and the awareness of flaws.

We observed a tendency for participants to conclude causation from correlation. In both studies, we found that the ratings for the awareness of ``correlation is not causation'' were positive (i.e., agreeing that the scatterplot implied causation) even when participants were warned that correlation is not causation. How do we reduce the illusion of causality when using question-answering systems? Here, we devise design considerations based on the study results.

\subsection{Encouraging Skepticism When Using Question-Answering Systems}

A core implication of our results is that question-answering systems could utilize visualizations of correlation to create an illusion of causality: By showing a scatterplot, these systems could increase the likelihood for users to accept causal claims that are unfounded. Mitigating the illusion of causality entails a deliberate design effort. In this section, we argue that encouraging users to be skeptical about automated answers could promote an appropriate interpretation of automatically generated causal claims and propose design ideas to inspire skepticism.

\begin{figure}[t!]
	\centering
	\includegraphics[width=\linewidth]{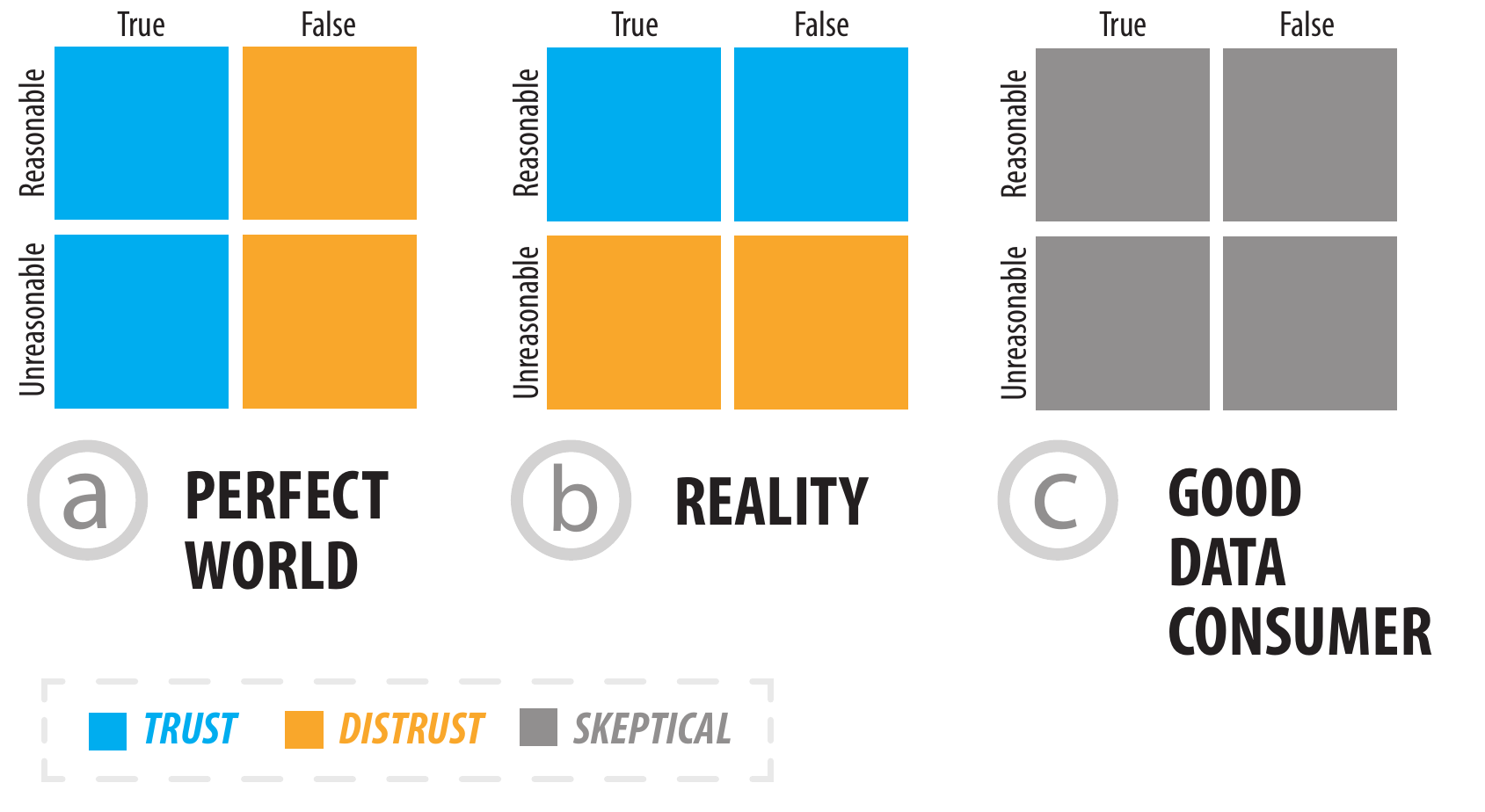}
	\caption{The relationship among trust, a claim's plausibility and the ground truth in different scenarios. In each bigger square, the y-axis is a claim's plausibility and the x-axis is the ground truth. In a perfect world (a), users should trust a claim only if it is true. In reality (b), users tend to trust a reasonable claim and distrust an unreasonable claim. When the truth is unknown, a good data consumer (c) should be skeptical despite a claims' plausibility.}
	\Description{Three illustrations showing how users should perceive causal claims in a perfect world, in reality, and as good data consumers. Read the caption for the details.}
	\label{skeptical}
\end{figure}

Why should users be skeptical when considering causal claims that are automatically generated? In a perfect world, user trust in a system's answer should match the ground truth---users should trust causal claims only when they are true and distrust false claims (Fig.~\ref{skeptical}a). In reality, however, belief in causal claims depends on their perceived plausibility despite the ground truth (Fig.~\ref{skeptical}b). Very often, users can only determine the plausibility of a causal claim but not whether it is true. Hence, we advocate that users should be skeptical whenever they cannot assess the veracity of a causal claim (Fig.~\ref{skeptical}c): A good data consumer should question the validity of a reasonable claim because the causal relationship could be fake; she should not refute the possibility of an unreasonable claim since the claim could hold true.

\subsubsection{Encouraging Skepticism for Reasonable Claims}

How do we encourage users to be skeptical about reasonable causal claims through interface design? Warning could be a potential solution. In study 1, we observed that participants tended to be more cautious in agreeing with a reasonable claim given a simple warning message. However, a simple warning could be unreliable: When the system only provided reasonable causal claims in study 2, the warning did not seem to promote such caution. To improve the effectiveness of warning in inspiring skepticism, its design could be improved based on research in warning science. For example, Wogalter~\cite{model} proposed the Communication-Human Information Processing (C-HIP) Model to describe the perceptual and cognitive processes after people see a warning. The model suggests asking a series of questions to assess the effectiveness of warning messages. For instance, do people notice the warning? Is the message in the warning being effectively communicated?

What are other interface design ideas to help encourage skepticism? In study 1, we found that participants tended to disagree less with unreasonable claims and agree more with hard-to-tell claims given a scatterplot. This indicates that correlation depicted in a scatterplot could induce an illusion of causality. To mitigate this illusion, it seems plausible to hide scatterplots from causal claims. Nevertheless, some participants felt that they trusted the system more because the scatterplots enabled them to see the data. Ideally, designers should keep the benefits of scatterplots while mitigating their side effects. Ritchie et al.~\cite{lie} found that transitioning from a non-deceptive view to a deceptive one could reduce the deception caused by the second view while enabling users to access the benefits of first view. Following this idea, a system could hide scatterplots by default while providing an option for users to view them. It would be interesting to investigate whether this design could reduce users' tendency to confuse correlation and causation.

How do we improve the general awareness of ``correlation is not causation''? Besides inspiring skepticism about a reasonable claim, warnings also appeared to raise awareness of ``correlation is not causation'' in study 1. Again, a simple warning alone could be ineffective: Even when participants were warned that correlation is not causation, they tended to agree that the scatterplot implied causation. This suggests that system developers might need to look beyond interface design to help users acquire correct statistical knowledge. Alternatives include pedagogical approaches such as tutorials. For example, when Tableau introduced Explain Data, they emphasized that users were the data experts, and they should judge the veracity of the causal claims based on their knowledge~\cite{tTutorial1, tTutorial2}. Future work will study the effectiveness of such tutorials in reducing users' tendency to associate correlation with causation.

\subsubsection{Encouraging Open-Mindedness for Unreasonable Claims}

While this work sheds light on ways to inspire skepticism for reasonable claims, designs to keep users open-minded when they see unreasonable claims are yet to be explored. Open-mindedness is a different form of skepticism: Instead of being skeptical about the automatically generated causal claims, users are skeptical about their beliefs and expectations about the data. Different models (e.g., Bayesian statistics~\cite{bayes} and the data-frame model~\cite{dataFrame1, dataFrame2}) have been developed to explain the process through which people update their beliefs. Prior research in misinformation showed that existing beliefs are rigid, and people are inclined to resist changes to their beliefs~\cite{misinfo}.

Although encouraging open-mindedness could be challenging, what are some potential ideas to keep people open-minded when they see unreasonable causal claims? An idea is to enable users to tell the system if an answer makes sense. If users consider an answer questionable, the system could explain why a causal relationship might exist to prevent users from prematurely rejecting a causal claim that seems unreasonable. However, further evidence would be required to demonstrate the effectiveness of this approach.

\subsection{Study Limitations and Future Work}

Our results hint at the potential for scatterplots to create an illusion of causality and the potential for a simple warning to reduce this illusion. We note that these are observations under controlled experiments, and we are prudent in drawing conclusions about the practical significance of the findings. First, to collect data from hundreds of people on MTurk, we needed to sacrifice realism to adapt the study for an online setting. For example, participants examined a series of answers provided by the system rather than really interact with a working system. Second, collaboration could protect users from being misled in practice: While an analyst might draw causal conclusions from correlational evidence, colleagues might remind the analyst of the flaw. Learning about the practical implications of our findings will require observing how people employ systems such as Explain Data~\cite{explainData} in their workflow and studying how people collaborate during data analysis.

We also note that the effectiveness of a simple warning in reducing causal illusion warrants further studies. Our work only compared four experimental conditions (Fig.~\ref{conditions}). Evidence from further comparisons (e.g., a comparison between an additional claim + vis + warning condition and the original claim + vis condition) could support our findings about the effectiveness of the warning. An ideal experiment is to consider each information type (claim, vis, textual description, and warning) an independent variable with two levels (with and without). This experiment will enable comparisons among all possible experimental conditions. Nevertheless, adding more conditions greatly reduces power given Bonferroni correction, and interesting findings might be missed. In future studies, experimenters would likely want to preserve power by honing in on a smaller set of comparisons. Our findings could provide guidance on what focused comparisons to make.

Our target population was potential end users of question-answering systems. These users include both people who are less proficient in data analysis and those who are more proficient. Our participants ranged from beginner users to more advanced analysts and appeared to be a reasonable proxy for our target. Yet, the focus on these users also implies that some findings (e.g., the tendency among participants to confuse correlation and causation) may not generalize if we conduct the studies with professional analysts only. Future work will replicate our study with these experts.

In both studies, we used a single question to measure trust in the question-answering system. In future, a questionnaire with multiple questions could be developed for assessing user trust in these systems. Such a questionnaire will measure sub-dimensions of trust (e.g., understanding) and enable researchers to learn about more fine-grained reasons for trusting a system (e.g., the system is trustworthy because users can easily understand the answers).

We have investigated whether showing correlational evidence could induce causal illusion. One form of correlational evidence we studied was textual description about correlation. We note that creating a description that completely eliminates casual perception could be challenging because people might easily mistake correlation for causation. Future research will investigate how different phrasing of correlation descriptions will affect casual perception.

Our study focused on numerical variables. Visualizing correlation between numerical variables using scatterplots is common. For other data types (e.g., categorical variables), other charts are often used. A natural extension to our work is to investigate the generalizability of our findings to other data types and charts.

Finally, although we are advocates for encouraging skepticism when using question-answering systems, we note that inspiring the right level of skepticism could be challenging. Ideally, users’ skepticism about a causal claim should match the evidence they have about the claim: For causal claims with good support (e.g., carbon dioxide emission leads to global warming), users could be less skeptical; for claims that lack supportive evidence, users could evaluate them more critically. However, it is difficult for a system to infer the amount of evidence users have about a claim and encourage skepticism accordingly. Moving forward, researchers could investigate whether telling users to be skeptical (e.g., through warnings) promotes an appropriate level of skepticism or engenders excessive and unhealthy skepticism.

\section{Conclusion}

Our work is situated in the discourse about the the ethical implications of data visualization~\cite{ethical}. We highlighted another scenario where visualizations might mislead users---question-answering systems could visualize correlation to create an illusion of causality. In particular, we found that in a system that occasionally provided unreasonable answers, showing a scatterplot next to a causal claim increased the plausibility of unreasonable and hard-to-tell claims. However, providing a simple warning about ``correlation is not causation'' seemed to lead participants to accept reasonable claims more cautiously. We further observed that our participants had a tendency to associate correlation with causation, but the warning appeared to reduce the tendency. We did not observe these effects of warning in a system that only provided reasonable answers. Based on the findings, we advocate that system developers could encourage users to be skeptical about answers generated by question-answering systems and have proposed ideas for doing so.

\begin{acks}
We thank Jian Zhao for early comments on the studies and the GT Visualization Lab for feedback on the paper. Special thanks to the anonymous reviewers for the thoughtful reviews.
\end{acks}

\bibliographystyle{ACM-Reference-Format}
\bibliography{sample-base}

%%% -*-BibTeX-*-
%%% Do NOT edit. File created by BibTeX with style
%%% ACM-Reference-Format-Journals [18-Jan-2012].

\begin{thebibliography}{72}

%%% ====================================================================
%%% NOTE TO THE USER: you can override these defaults by providing
%%% customized versions of any of these macros before the \bibliography
%%% command.  Each of them MUST provide its own final punctuation,
%%% except for \shownote{}, \showDOI{}, and \showURL{}.  The latter two
%%% do not use final punctuation, in order to avoid confusing it with
%%% the Web address.
%%%
%%% To suppress output of a particular field, define its macro to expand
%%% to an empty string, or better, \unskip, like this:
%%%
%%% \newcommand{\showDOI}[1]{\unskip}   % LaTeX syntax
%%%
%%% \def \showDOI #1{\unskip}           % plain TeX syntax
%%%
%%% ====================================================================

\ifx \showCODEN    \undefined \def \showCODEN     #1{\unskip}     \fi
\ifx \showDOI      \undefined \def \showDOI       #1{#1}\fi
\ifx \showISBNx    \undefined \def \showISBNx     #1{\unskip}     \fi
\ifx \showISBNxiii \undefined \def \showISBNxiii  #1{\unskip}     \fi
\ifx \showISSN     \undefined \def \showISSN      #1{\unskip}     \fi
\ifx \showLCCN     \undefined \def \showLCCN      #1{\unskip}     \fi
\ifx \shownote     \undefined \def \shownote      #1{#1}          \fi
\ifx \showarticletitle \undefined \def \showarticletitle #1{#1}   \fi
\ifx \showURL      \undefined \def \showURL       {\relax}        \fi
% The following commands are used for tagged output and should be
% invisible to TeX
\providecommand\bibfield[2]{#2}
\providecommand\bibinfo[2]{#2}
\providecommand\natexlab[1]{#1}
\providecommand\showeprint[2][]{arXiv:#2}

\bibitem[\protect\citeauthoryear{AI}{AI}{2017}]%
        {business}
\bibfield{author}{\bibinfo{person}{Outlier AI}.}
  \bibinfo{year}{2017}\natexlab{}.
\newblock \bibinfo{booktitle}{\emph{How to Conduct a Proper Root Cause
  Analysis}}.
\newblock
\urldef\tempurl%
\url{https://towardsdatascience.com/how-to-conduct-a-proper-root-cause-analysis-789b9847f84b}
\showURL{%
\tempurl}


\bibitem[\protect\citeauthoryear{Amar, Eagan, and Stasko}{Amar
  et~al\mbox{.}}{2005}]%
        {lowLevel}
\bibfield{author}{\bibinfo{person}{Robert Amar}, \bibinfo{person}{James Eagan},
  {and} \bibinfo{person}{John Stasko}.} \bibinfo{year}{2005}\natexlab{}.
\newblock \showarticletitle{Low-Level Components of Analytic Activity in
  Information Visualization}. In \bibinfo{booktitle}{\emph{IEEE Symposium on
  Information Visualization}}. \bibinfo{publisher}{IEEE},
  \bibinfo{pages}{111--117}.
\newblock
\urldef\tempurl%
\url{https://doi.org/10.1109/INFVIS.2005.1532136}
\showDOI{\tempurl}


\bibitem[\protect\citeauthoryear{Apple}{Apple}{2020}]%
        {siri}
\bibfield{author}{\bibinfo{person}{Apple}.} \bibinfo{year}{2020}\natexlab{}.
\newblock \bibinfo{title}{Siri - Apple}.
\newblock
\newblock
\urldef\tempurl%
\url{https://www.apple.com/siri/}
\showURL{%
\tempurl}


\bibitem[\protect\citeauthoryear{Ashktorab, Jain, Liao, and Weisz}{Ashktorab
  et~al\mbox{.}}{2019}]%
        {breakdown}
\bibfield{author}{\bibinfo{person}{Zahra Ashktorab}, \bibinfo{person}{Mohit
  Jain}, \bibinfo{person}{Q~Vera Liao}, {and} \bibinfo{person}{Justin~D
  Weisz}.} \bibinfo{year}{2019}\natexlab{}.
\newblock \showarticletitle{Resilient Chatbots: Repair Strategy Preferences for
  Conversational Breakdowns}. In \bibinfo{booktitle}{\emph{Proceedings of the
  2019 CHI Conference on Human Factors in Computing Systems}}.
  \bibinfo{publisher}{ACM}, \bibinfo{pages}{1--12}.
\newblock
\urldef\tempurl%
\url{https://doi.org/10.1145/3290605.3300484}
\showDOI{\tempurl}


\bibitem[\protect\citeauthoryear{Bolstad and Curran}{Bolstad and
  Curran}{2016}]%
        {bayes}
\bibfield{author}{\bibinfo{person}{William~M Bolstad} {and}
  \bibinfo{person}{James~M Curran}.} \bibinfo{year}{2016}\natexlab{}.
\newblock \bibinfo{booktitle}{\emph{Introduction to Bayesian Statistics}}.
\newblock \bibinfo{publisher}{John Wiley \& Sons}.
\newblock
\showISBNx{978-1-118-09156-2}


\bibitem[\protect\citeauthoryear{Borkin, Bylinskii, Kim, Bainbridge, Yeh,
  Borkin, Pfister, and Oliva}{Borkin et~al\mbox{.}}{2015}]%
        {memorability}
\bibfield{author}{\bibinfo{person}{Michelle~A Borkin}, \bibinfo{person}{Zoya
  Bylinskii}, \bibinfo{person}{Nam~Wook Kim}, \bibinfo{person}{Constance~May
  Bainbridge}, \bibinfo{person}{Chelsea~S Yeh}, \bibinfo{person}{Daniel
  Borkin}, \bibinfo{person}{Hanspeter Pfister}, {and} \bibinfo{person}{Aude
  Oliva}.} \bibinfo{year}{2015}\natexlab{}.
\newblock \showarticletitle{Beyond Memorability: Visualization Recognition and
  Recall}.
\newblock \bibinfo{journal}{\emph{IEEE Transactions on Visualization and
  Computer Graphics}} \bibinfo{volume}{22}, \bibinfo{number}{1}
  (\bibinfo{year}{2015}), \bibinfo{pages}{519--528}.
\newblock
\urldef\tempurl%
\url{https://doi.org/10.1109/TVCG.2015.2467732}
\showDOI{\tempurl}


\bibitem[\protect\citeauthoryear{Boy, Pandey, Emerson, Satterthwaite, Nov, and
  Bertini}{Boy et~al\mbox{.}}{2017}]%
        {icons}
\bibfield{author}{\bibinfo{person}{Jeremy Boy}, \bibinfo{person}{Anshul~Vikram
  Pandey}, \bibinfo{person}{John Emerson}, \bibinfo{person}{Margaret
  Satterthwaite}, \bibinfo{person}{Oded Nov}, {and} \bibinfo{person}{Enrico
  Bertini}.} \bibinfo{year}{2017}\natexlab{}.
\newblock \showarticletitle{Showing People Behind Data: Does Anthropomorphizing
  Visualizations Elicit More Empathy for Human Rights Data?}. In
  \bibinfo{booktitle}{\emph{Proceedings of the 2017 CHI Conference on Human
  Factors in Computing Systems}}. \bibinfo{publisher}{ACM},
  \bibinfo{pages}{5462--5474}.
\newblock
\urldef\tempurl%
\url{https://doi.org/10.1145/3025453.3025512}
\showDOI{\tempurl}


\bibitem[\protect\citeauthoryear{Bureau}{Bureau}{2020}]%
        {census}
\bibfield{author}{\bibinfo{person}{United States~Census Bureau}.}
  \bibinfo{year}{2020}\natexlab{}.
\newblock \bibinfo{title}{Census Bureau}.
\newblock
\newblock
\urldef\tempurl%
\url{https://www.census.gov}
\showURL{%
\tempurl}


\bibitem[\protect\citeauthoryear{Cheng, Wang, Zhang, O'Connell, Gray, Harper,
  and Zhu}{Cheng et~al\mbox{.}}{2019}]%
        {ml}
\bibfield{author}{\bibinfo{person}{Hao-Fei Cheng}, \bibinfo{person}{Ruotong
  Wang}, \bibinfo{person}{Zheng Zhang}, \bibinfo{person}{Fiona O'Connell},
  \bibinfo{person}{Terrance Gray}, \bibinfo{person}{F~Maxwell Harper}, {and}
  \bibinfo{person}{Haiyi Zhu}.} \bibinfo{year}{2019}\natexlab{}.
\newblock \showarticletitle{Explaining Decision-Making Algorithms Through UI:
  Strategies to Help Non-Expert Stakeholders}. In
  \bibinfo{booktitle}{\emph{Proceedings of the 2019 CHI Conference on Human
  Factors in Computing Systems}}. \bibinfo{publisher}{ACM},
  \bibinfo{pages}{1--12}.
\newblock
\urldef\tempurl%
\url{https://doi.org/10.1145/3290605.3300789}
\showDOI{\tempurl}


\bibitem[\protect\citeauthoryear{Cleveland, Diaconis, and McGill}{Cleveland
  et~al\mbox{.}}{1982}]%
        {cleveland}
\bibfield{author}{\bibinfo{person}{William~S Cleveland}, \bibinfo{person}{Persi
  Diaconis}, {and} \bibinfo{person}{Robert McGill}.}
  \bibinfo{year}{1982}\natexlab{}.
\newblock \showarticletitle{Variables on Scatterplots Look More Highly
  Correlated When the Scales Are Increased}.
\newblock \bibinfo{journal}{\emph{Science}} \bibinfo{volume}{216},
  \bibinfo{number}{4550} (\bibinfo{year}{1982}), \bibinfo{pages}{1138--1141}.
\newblock
\urldef\tempurl%
\url{https://doi.org/10.1126/science.216.4550.1138}
\showDOI{\tempurl}


\bibitem[\protect\citeauthoryear{College}{College}{2020}]%
        {rca2}
\bibfield{author}{\bibinfo{person}{Wharton County~Junior College}.}
  \bibinfo{year}{2020}\natexlab{}.
\newblock \bibinfo{title}{Root Cause Analysis Training}.
\newblock
\newblock
\urldef\tempurl%
\url{https://www.wcjc.edu/Programs/continuing-education/root-cause.aspx}
\showURL{%
\tempurl}


\bibitem[\protect\citeauthoryear{Correll}{Correll}{2019}]%
        {ethical}
\bibfield{author}{\bibinfo{person}{Michael Correll}.}
  \bibinfo{year}{2019}\natexlab{}.
\newblock \showarticletitle{Ethical Dimensions of Visualization Research}. In
  \bibinfo{booktitle}{\emph{Proceedings of the 2019 CHI Conference on Human
  Factors in Computing Systems}}. \bibinfo{publisher}{ACM},
  \bibinfo{pages}{1--13}.
\newblock
\urldef\tempurl%
\url{https://doi.org/10.1145/3290605.3300418}
\showDOI{\tempurl}


\bibitem[\protect\citeauthoryear{Correll, Bertini, and Franconeri}{Correll
  et~al\mbox{.}}{2020}]%
        {truncate}
\bibfield{author}{\bibinfo{person}{Michael Correll}, \bibinfo{person}{Enrico
  Bertini}, {and} \bibinfo{person}{Steven Franconeri}.}
  \bibinfo{year}{2020}\natexlab{}.
\newblock \showarticletitle{Truncating the Y-Axis: Threat or Menace?}. In
  \bibinfo{booktitle}{\emph{Proceedings of the 2020 CHI Conference on Human
  Factors in Computing Systems}}. \bibinfo{publisher}{ACM},
  \bibinfo{pages}{1--12}.
\newblock
\urldef\tempurl%
\url{https://doi.org/10.1145/3313831.3376222}
\showDOI{\tempurl}


\bibitem[\protect\citeauthoryear{Correll and Heer}{Correll and Heer}{2017}]%
        {blackhat}
\bibfield{author}{\bibinfo{person}{Michael Correll} {and}
  \bibinfo{person}{Jeffrey Heer}.} \bibinfo{year}{2017}\natexlab{}.
\newblock \showarticletitle{Black Hat Visualization}. In
  \bibinfo{booktitle}{\emph{Workshop on Dealing with Cognitive Biases in
  Visualisations (DECISIVe)}}.
\newblock
\urldef\tempurl%
\url{https://decisive-workshop.dbvis.de/wp-content/uploads/2017/09/0115-paper.pdf}
\showURL{%
\tempurl}


\bibitem[\protect\citeauthoryear{Coursera}{Coursera}{2020}]%
        {rca1}
\bibfield{author}{\bibinfo{person}{Coursera}.} \bibinfo{year}{2020}\natexlab{}.
\newblock \bibinfo{title}{Root Cause Analysis - Root Cause Analysis $|$
  Coursera}.
\newblock
\newblock
\urldef\tempurl%
\url{https://www.coursera.org/lecture/six-sigma-analyze/root-cause-analysis-w01Qj}
\showURL{%
\tempurl}


\bibitem[\protect\citeauthoryear{Dhamdhere, McCurley, Nahmias, Sundararajan,
  and Yan}{Dhamdhere et~al\mbox{.}}{2017}]%
        {analyza}
\bibfield{author}{\bibinfo{person}{Kedar Dhamdhere}, \bibinfo{person}{Kevin~S
  McCurley}, \bibinfo{person}{Ralfi Nahmias}, \bibinfo{person}{Mukund
  Sundararajan}, {and} \bibinfo{person}{Qiqi Yan}.}
  \bibinfo{year}{2017}\natexlab{}.
\newblock \showarticletitle{Analyza: Exploring Data With Conversation}. In
  \bibinfo{booktitle}{\emph{Proceedings of the 22nd International Conference on
  Intelligent User Interfaces}}. \bibinfo{publisher}{ACM},
  \bibinfo{pages}{493--504}.
\newblock
\urldef\tempurl%
\url{https://doi.org/10.1145/3025171.3025227}
\showDOI{\tempurl}


\bibitem[\protect\citeauthoryear{Dimara, Bailly, Bezerianos, and
  Franconeri}{Dimara et~al\mbox{.}}{2018}]%
        {attraction}
\bibfield{author}{\bibinfo{person}{Evanthia Dimara}, \bibinfo{person}{Gilles
  Bailly}, \bibinfo{person}{Anastasia Bezerianos}, {and}
  \bibinfo{person}{Steven Franconeri}.} \bibinfo{year}{2018}\natexlab{}.
\newblock \showarticletitle{Mitigating the Attraction Effect With
  Visualizations}.
\newblock \bibinfo{journal}{\emph{IEEE Transactions on Visualization and
  Computer Graphics}} \bibinfo{volume}{25}, \bibinfo{number}{1}
  (\bibinfo{year}{2018}), \bibinfo{pages}{850--860}.
\newblock
\urldef\tempurl%
\url{https://doi.org/10.1109/TVCG.2018.2865233}
\showDOI{\tempurl}


\bibitem[\protect\citeauthoryear{Dimara, Franconeri, Plaisant, Bezerianos, and
  Dragicevic}{Dimara et~al\mbox{.}}{2020}]%
        {cognitive}
\bibfield{author}{\bibinfo{person}{Evanthia Dimara}, \bibinfo{person}{Steven
  Franconeri}, \bibinfo{person}{Catherine Plaisant}, \bibinfo{person}{Anastasia
  Bezerianos}, {and} \bibinfo{person}{Pierre Dragicevic}.}
  \bibinfo{year}{2020}\natexlab{}.
\newblock \showarticletitle{A task-based taxonomy of cognitive biases for
  information visualization}.
\newblock \bibinfo{journal}{\emph{IEEE Transactions on Visualization and
  Computer Graphics}} \bibinfo{volume}{26}, \bibinfo{number}{2}
  (\bibinfo{year}{2020}), \bibinfo{pages}{1413--1432}.
\newblock
\urldef\tempurl%
\url{https://doi.org/10.1109/TVCG.2018.2872577}
\showDOI{\tempurl}


\bibitem[\protect\citeauthoryear{Egelman, Cranor, and Hong}{Egelman
  et~al\mbox{.}}{2008}]%
        {phishing}
\bibfield{author}{\bibinfo{person}{Serge Egelman},
  \bibinfo{person}{Lorrie~Faith Cranor}, {and} \bibinfo{person}{Jason Hong}.}
  \bibinfo{year}{2008}\natexlab{}.
\newblock \showarticletitle{You've Been Warned: An Empirical Study of the
  Effectiveness of Web Browser Phishing Warnings}. In
  \bibinfo{booktitle}{\emph{Proceedings of the 2008 CHI Conference on Human
  Factors in Computing Systems}}. \bibinfo{publisher}{ACM},
  \bibinfo{pages}{1065--1074}.
\newblock
\urldef\tempurl%
\url{https://doi.org/10.1145/1357054.1357219}
\showDOI{\tempurl}


\bibitem[\protect\citeauthoryear{Fast, Chen, Mendelsohn, Bassen, and
  Bernstein}{Fast et~al\mbox{.}}{2018}]%
        {dataSci}
\bibfield{author}{\bibinfo{person}{Ethan Fast}, \bibinfo{person}{Binbin Chen},
  \bibinfo{person}{Julia Mendelsohn}, \bibinfo{person}{Jonathan Bassen}, {and}
  \bibinfo{person}{Michael~S Bernstein}.} \bibinfo{year}{2018}\natexlab{}.
\newblock \showarticletitle{Iris: A Conversational Agent for Complex Tasks}. In
  \bibinfo{booktitle}{\emph{Proceedings of the 2018 CHI Conference on Human
  Factors in Computing Systems}}. \bibinfo{publisher}{ACM},
  \bibinfo{pages}{1--13}.
\newblock
\urldef\tempurl%
\url{https://doi.org/10.1145/3173574.3174047}
\showDOI{\tempurl}


\bibitem[\protect\citeauthoryear{F{\o}lstad and Brandtz{\ae}g}{F{\o}lstad and
  Brandtz{\ae}g}{2017}]%
        {NLOpinion}
\bibfield{author}{\bibinfo{person}{Asbj{\o}rn F{\o}lstad} {and}
  \bibinfo{person}{Petter~Bae Brandtz{\ae}g}.} \bibinfo{year}{2017}\natexlab{}.
\newblock \showarticletitle{Chatbots and the New World of HCI}.
\newblock \bibinfo{journal}{\emph{Interactions}} \bibinfo{volume}{24},
  \bibinfo{number}{4} (\bibinfo{year}{2017}), \bibinfo{pages}{38--42}.
\newblock
\urldef\tempurl%
\url{https://doi.org/10.1145/3085558}
\showDOI{\tempurl}


\bibitem[\protect\citeauthoryear{for Education~Statistics}{for
  Education~Statistics}{2020}]%
        {nces}
\bibfield{author}{\bibinfo{person}{National~Center for Education~Statistics}.}
  \bibinfo{year}{2020}\natexlab{}.
\newblock \bibinfo{title}{National Center for Education Statistics (NCES) Home
  Page, a part of the U.S. Department of Education}.
\newblock
\newblock
\urldef\tempurl%
\url{https://nces.ed.gov}
\showURL{%
\tempurl}


\bibitem[\protect\citeauthoryear{Foundation}{Foundation}{2020}]%
        {kff}
\bibfield{author}{\bibinfo{person}{Kaiser~Family Foundation}.}
  \bibinfo{year}{2020}\natexlab{}.
\newblock \bibinfo{title}{KFF - Health Policy Analysis, Polling and
  Journalism}.
\newblock
\newblock
\urldef\tempurl%
\url{https://www.kff.org}
\showURL{%
\tempurl}


\bibitem[\protect\citeauthoryear{Gao, Dontcheva, Adar, Liu, and Karahalios}{Gao
  et~al\mbox{.}}{2015}]%
        {datatone}
\bibfield{author}{\bibinfo{person}{Tong Gao}, \bibinfo{person}{Mira Dontcheva},
  \bibinfo{person}{Eytan Adar}, \bibinfo{person}{Zhicheng Liu}, {and}
  \bibinfo{person}{Karrie~G Karahalios}.} \bibinfo{year}{2015}\natexlab{}.
\newblock \showarticletitle{Datatone: Managing Ambiguity in Natural Language
  Interfaces for Data Visualization}. In \bibinfo{booktitle}{\emph{Proceedings
  of the 28th Annual ACM Symposium on User Interface Software \& Technology}}.
  \bibinfo{publisher}{ACM}, \bibinfo{pages}{489--500}.
\newblock
\urldef\tempurl%
\url{https://doi.org/10.1145/2807442.2807478}
\showDOI{\tempurl}


\bibitem[\protect\citeauthoryear{Hearst and Tory}{Hearst and Tory}{2019}]%
        {visualConverse}
\bibfield{author}{\bibinfo{person}{Marti Hearst} {and} \bibinfo{person}{Melanie
  Tory}.} \bibinfo{year}{2019}\natexlab{}.
\newblock \showarticletitle{Would You Like a Chart With That? Incorporating
  Visualizations Into Conversational Interfaces}. In
  \bibinfo{booktitle}{\emph{2019 IEEE Visualization Conference (VIS)}}. IEEE,
  \bibinfo{pages}{1--5}.
\newblock
\urldef\tempurl%
\url{https://doi.org/10.1109/VISUAL.2019.89337668}
\showDOI{\tempurl}


\bibitem[\protect\citeauthoryear{Hearst, Tory, and Setlur}{Hearst
  et~al\mbox{.}}{2019}]%
        {vagueMod}
\bibfield{author}{\bibinfo{person}{Marti Hearst}, \bibinfo{person}{Melanie
  Tory}, {and} \bibinfo{person}{Vidya Setlur}.}
  \bibinfo{year}{2019}\natexlab{}.
\newblock \showarticletitle{Toward Interface Defaults for Vague Modifiers in
  Natural Language Interfaces for Visual Analysis}. In
  \bibinfo{booktitle}{\emph{2019 IEEE Visualization Conference (VIS)}}. IEEE,
  \bibinfo{pages}{21--25}.
\newblock
\urldef\tempurl%
\url{https://doi.org/10.1109/VISUAL.2019.8933569}
\showDOI{\tempurl}


\bibitem[\protect\citeauthoryear{Heer and Agrawala}{Heer and Agrawala}{2006}]%
        {lineChart}
\bibfield{author}{\bibinfo{person}{Jeffrey Heer} {and} \bibinfo{person}{Maneesh
  Agrawala}.} \bibinfo{year}{2006}\natexlab{}.
\newblock \showarticletitle{Multi-Scale Banking to 45 Degrees}.
\newblock \bibinfo{journal}{\emph{IEEE Transactions on Visualization and
  Computer Graphics}} \bibinfo{volume}{12}, \bibinfo{number}{5}
  (\bibinfo{year}{2006}), \bibinfo{pages}{701--708}.
\newblock
\urldef\tempurl%
\url{https://doi.org/10.1109/TVCG.2006.163}
\showDOI{\tempurl}


\bibitem[\protect\citeauthoryear{Hellier, Wright, Edworthy, and
  Newstead}{Hellier et~al\mbox{.}}{2000}]%
        {arousal}
\bibfield{author}{\bibinfo{person}{Elizabeth Hellier},
  \bibinfo{person}{Daniel~B Wright}, \bibinfo{person}{Judy Edworthy}, {and}
  \bibinfo{person}{Stephen Newstead}.} \bibinfo{year}{2000}\natexlab{}.
\newblock \showarticletitle{On the Stability of the Arousal Strength of Warning
  Signal Words}.
\newblock \bibinfo{journal}{\emph{Applied Cognitive Psychology: The Official
  Journal of the Society for Applied Research in Memory and Cognition}}
  \bibinfo{volume}{14}, \bibinfo{number}{6} (\bibinfo{year}{2000}),
  \bibinfo{pages}{577--592}.
\newblock
\urldef\tempurl%
\url{https://doi.org/10.1002/1099-0720(200011/12)14:6<577::AID-ACP682>3.0.CO;2-A}
\showDOI{\tempurl}


\bibitem[\protect\citeauthoryear{Herlocker, Konstan, and Riedl}{Herlocker
  et~al\mbox{.}}{2000}]%
        {cfs}
\bibfield{author}{\bibinfo{person}{Jonathan~L Herlocker},
  \bibinfo{person}{Joseph~A Konstan}, {and} \bibinfo{person}{John Riedl}.}
  \bibinfo{year}{2000}\natexlab{}.
\newblock \showarticletitle{Explaining Collaborative Filtering
  Recommendations}. In \bibinfo{booktitle}{\emph{Proceedings of the 2000 ACM
  Conference on Computer Supported Cooperative Work}}.
  \bibinfo{publisher}{ACM}, \bibinfo{pages}{241--250}.
\newblock
\urldef\tempurl%
\url{https://doi.org/10.1145/358916.358995}
\showDOI{\tempurl}


\bibitem[\protect\citeauthoryear{Hoque, Setlur, Tory, and Dykeman}{Hoque
  et~al\mbox{.}}{2017}]%
        {evizeon}
\bibfield{author}{\bibinfo{person}{Enamul Hoque}, \bibinfo{person}{Vidya
  Setlur}, \bibinfo{person}{Melanie Tory}, {and} \bibinfo{person}{Isaac
  Dykeman}.} \bibinfo{year}{2017}\natexlab{}.
\newblock \showarticletitle{Applying Pragmatics Principles for Interaction With
  Visual Analytics}.
\newblock \bibinfo{journal}{\emph{IEEE Transactions on Visualization and
  Computer Graphics}} \bibinfo{volume}{24}, \bibinfo{number}{1}
  (\bibinfo{year}{2017}), \bibinfo{pages}{309--318}.
\newblock
\urldef\tempurl%
\url{https://doi.org/10.1109/TVCG.2017.2744684}
\showDOI{\tempurl}


\bibitem[\protect\citeauthoryear{Hullman and Diakopoulos}{Hullman and
  Diakopoulos}{2011}]%
        {rhetoric}
\bibfield{author}{\bibinfo{person}{Jessica Hullman} {and} \bibinfo{person}{Nick
  Diakopoulos}.} \bibinfo{year}{2011}\natexlab{}.
\newblock \showarticletitle{Visualization Rhetoric: Framing Effects in
  Narrative Visualization}.
\newblock \bibinfo{journal}{\emph{IEEE Transactions on Visualization and
  Computer Graphics}} \bibinfo{volume}{17}, \bibinfo{number}{12}
  (\bibinfo{year}{2011}), \bibinfo{pages}{2231--2240}.
\newblock
\urldef\tempurl%
\url{https://doi.org/10.1109/TVCG.2011.255}
\showDOI{\tempurl}


\bibitem[\protect\citeauthoryear{Kim, Reinecke, and Hullman}{Kim
  et~al\mbox{.}}{2017}]%
        {otherEyes}
\bibfield{author}{\bibinfo{person}{Yea-Seul Kim}, \bibinfo{person}{Katharina
  Reinecke}, {and} \bibinfo{person}{Jessica Hullman}.}
  \bibinfo{year}{2017}\natexlab{}.
\newblock \showarticletitle{Data Through Others' Eyes: The Impact of
  Visualizing Others' Expectations on Visualization Interpretation}.
\newblock \bibinfo{journal}{\emph{IEEE Transactions on Visualization and
  Computer Graphics}} \bibinfo{volume}{24}, \bibinfo{number}{1}
  (\bibinfo{year}{2017}), \bibinfo{pages}{760--769}.
\newblock
\urldef\tempurl%
\url{https://doi.org/10.1109/TVCG.2017.2745240}
\showDOI{\tempurl}


\bibitem[\protect\citeauthoryear{Klein, Moon, and Hoffman}{Klein
  et~al\mbox{.}}{2006a}]%
        {dataFrame1}
\bibfield{author}{\bibinfo{person}{Gary Klein}, \bibinfo{person}{Brian Moon},
  {and} \bibinfo{person}{Robert~R Hoffman}.} \bibinfo{year}{2006}\natexlab{a}.
\newblock \showarticletitle{Making Sense of Sensemaking 1: Alternative
  Perspectives}.
\newblock \bibinfo{journal}{\emph{IEEE Intelligent Systems}}
  \bibinfo{volume}{21}, \bibinfo{number}{4} (\bibinfo{year}{2006}),
  \bibinfo{pages}{70--73}.
\newblock
\urldef\tempurl%
\url{https://doi.org/10.1109/MIS.2006.75}
\showDOI{\tempurl}


\bibitem[\protect\citeauthoryear{Klein, Moon, and Hoffman}{Klein
  et~al\mbox{.}}{2006b}]%
        {dataFrame2}
\bibfield{author}{\bibinfo{person}{Gary Klein}, \bibinfo{person}{Brian Moon},
  {and} \bibinfo{person}{Robert~R Hoffman}.} \bibinfo{year}{2006}\natexlab{b}.
\newblock \showarticletitle{Making Sense of Sensemaking 2: A Macrocognitive
  Model}.
\newblock \bibinfo{journal}{\emph{IEEE Intelligent Systems}}
  \bibinfo{volume}{21}, \bibinfo{number}{5} (\bibinfo{year}{2006}),
  \bibinfo{pages}{88--92}.
\newblock
\urldef\tempurl%
\url{https://doi.org/10.1109/MIS.2006.100}
\showDOI{\tempurl}


\bibitem[\protect\citeauthoryear{Kong, Liu, and Karahalios}{Kong
  et~al\mbox{.}}{2018}]%
        {frame1}
\bibfield{author}{\bibinfo{person}{Ha-Kyung Kong}, \bibinfo{person}{Zhicheng
  Liu}, {and} \bibinfo{person}{Karrie Karahalios}.}
  \bibinfo{year}{2018}\natexlab{}.
\newblock \showarticletitle{Frames and Slants in Titles of Visualizations on
  Controversial Topics}. In \bibinfo{booktitle}{\emph{Proceedings of the 2018
  CHI Conference on Human Factors in Computing Systems}}.
  \bibinfo{publisher}{ACM}, \bibinfo{pages}{1--12}.
\newblock
\urldef\tempurl%
\url{https://doi.org/10.1145/3173574.3174012}
\showDOI{\tempurl}


\bibitem[\protect\citeauthoryear{Kong, Liu, and Karahalios}{Kong
  et~al\mbox{.}}{2019}]%
        {frame2}
\bibfield{author}{\bibinfo{person}{Ha-Kyung Kong}, \bibinfo{person}{Zhicheng
  Liu}, {and} \bibinfo{person}{Karrie Karahalios}.}
  \bibinfo{year}{2019}\natexlab{}.
\newblock \showarticletitle{Trust and Recall of Information Across Varying
  Degrees of Title-Visualization Misalignment}. In
  \bibinfo{booktitle}{\emph{Proceedings of the 2019 CHI Conference on Human
  Factors in Computing Systems}}. \bibinfo{publisher}{ACM},
  \bibinfo{pages}{1--13}.
\newblock
\urldef\tempurl%
\url{https://doi.org/10.1145/3290605.3300576}
\showDOI{\tempurl}


\bibitem[\protect\citeauthoryear{Law, Basole, and Wu}{Law
  et~al\mbox{.}}{2018}]%
        {me2}
\bibfield{author}{\bibinfo{person}{Po-Ming Law}, \bibinfo{person}{Rahul~C
  Basole}, {and} \bibinfo{person}{Yanhong Wu}.}
  \bibinfo{year}{2018}\natexlab{}.
\newblock \showarticletitle{Duet: Helping Data Analysis Novices Conduct
  Pairwise Comparisons by Minimal Specification}.
\newblock \bibinfo{journal}{\emph{IEEE Transactions on Visualization and
  Computer Graphics}} \bibinfo{volume}{25}, \bibinfo{number}{1}
  (\bibinfo{year}{2018}), \bibinfo{pages}{427--437}.
\newblock
\urldef\tempurl%
\url{https://doi.org/10.1109/TVCG.2018.2864526}
\showDOI{\tempurl}


\bibitem[\protect\citeauthoryear{Law, Das, and Basole}{Law
  et~al\mbox{.}}{2019}]%
        {me1}
\bibfield{author}{\bibinfo{person}{Po-Ming Law}, \bibinfo{person}{Subhajit
  Das}, {and} \bibinfo{person}{Rahul~C Basole}.}
  \bibinfo{year}{2019}\natexlab{}.
\newblock \showarticletitle{Comparing Apples and Oranges: Taxonomy and Design
  of Pairwise Comparisons within Tabular Data}. In
  \bibinfo{booktitle}{\emph{Proceedings of the 2019 CHI Conference on Human
  Factors in Computing Systems}}. \bibinfo{publisher}{ACM},
  \bibinfo{pages}{1--12}.
\newblock
\urldef\tempurl%
\url{https://doi.org/10.1145/3290605.3300409}
\showDOI{\tempurl}


\bibitem[\protect\citeauthoryear{Law, Endert, and Stasko}{Law
  et~al\mbox{.}}{2020}]%
        {characterizing}
\bibfield{author}{\bibinfo{person}{Po-Ming Law}, \bibinfo{person}{Alex Endert},
  {and} \bibinfo{person}{John Stasko}.} \bibinfo{year}{2020}\natexlab{}.
\newblock \showarticletitle{Characterizing Automated Data Insights}.
\newblock \bibinfo{journal}{\emph{arXiv preprint arXiv:2008.13060}}
  (\bibinfo{year}{2020}).
\newblock
\urldef\tempurl%
\url{https://arxiv.org/abs/2008.13060}
\showURL{%
\tempurl}


\bibitem[\protect\citeauthoryear{Lee}{Lee}{2020}]%
        {medium1}
\bibfield{author}{\bibinfo{person}{Victoria Lee}.}
  \bibinfo{year}{2020}\natexlab{}.
\newblock \bibinfo{booktitle}{\emph{Why We Quarantine: A Data Driven Love
  Letter to You and the Loves of Your Life}}.
\newblock
\urldef\tempurl%
\url{https://medium.com/swlh/why-we-quarantine-a-data-driven-love-letter-to-you-and-the-loves-of-your-life-c19de2bca87f}
\showURL{%
\tempurl}


\bibitem[\protect\citeauthoryear{Lehmann}{Lehmann}{2006}]%
        {nonparametric}
\bibfield{author}{\bibinfo{person}{Erich~L Lehmann}.}
  \bibinfo{year}{2006}\natexlab{}.
\newblock \bibinfo{booktitle}{\emph{Nonparametrics: Statistical Methods Based
  on Ranks}}.
\newblock \bibinfo{publisher}{Springer-Verlag}, \bibinfo{address}{New York, NY,
  USA}.
\newblock
\showISBNx{978-0-387-35212-1}


\bibitem[\protect\citeauthoryear{Lewandowsky, Ecker, Seifert, Schwarz, and
  Cook}{Lewandowsky et~al\mbox{.}}{2012}]%
        {misinfo}
\bibfield{author}{\bibinfo{person}{Stephan Lewandowsky},
  \bibinfo{person}{Ullrich~KH Ecker}, \bibinfo{person}{Colleen~M Seifert},
  \bibinfo{person}{Norbert Schwarz}, {and} \bibinfo{person}{John Cook}.}
  \bibinfo{year}{2012}\natexlab{}.
\newblock \showarticletitle{Misinformation and Its Correction: Continued
  Influence and Successful Debiasing}.
\newblock \bibinfo{journal}{\emph{Psychological Science in the Public
  Interest}} \bibinfo{volume}{13}, \bibinfo{number}{3} (\bibinfo{year}{2012}),
  \bibinfo{pages}{106--131}.
\newblock
\urldef\tempurl%
\url{https://doi.org/10.1177/1529100612451018}
\showDOI{\tempurl}


\bibitem[\protect\citeauthoryear{Liao, Davis, Geyer, Muller, and Shami}{Liao
  et~al\mbox{.}}{2016}]%
        {social}
\bibfield{author}{\bibinfo{person}{Q~Vera Liao}, \bibinfo{person}{Matthew
  Davis}, \bibinfo{person}{Werner Geyer}, \bibinfo{person}{Michael Muller},
  {and} \bibinfo{person}{N~Sadat Shami}.} \bibinfo{year}{2016}\natexlab{}.
\newblock \showarticletitle{What Can You Do? Studying Social-Agent Orientation
  and Agent Proactive Interactions With an Agent for Employees}. In
  \bibinfo{booktitle}{\emph{Proceedings of the 2016 ACM Conference on Designing
  Interactive Systems}}. \bibinfo{publisher}{ACM}, \bibinfo{pages}{264--275}.
\newblock
\urldef\tempurl%
\url{https://doi.org/10.1145/2901790.2901842}
\showDOI{\tempurl}


\bibitem[\protect\citeauthoryear{Liao, Mas-ud Hussain, Chandar, Davis,
  Khazaeni, Crasso, Wang, Muller, Shami, and Geyer}{Liao et~al\mbox{.}}{2018}]%
        {work}
\bibfield{author}{\bibinfo{person}{Q~Vera Liao}, \bibinfo{person}{Muhammed
  Mas-ud Hussain}, \bibinfo{person}{Praveen Chandar}, \bibinfo{person}{Matthew
  Davis}, \bibinfo{person}{Yasaman Khazaeni}, \bibinfo{person}{Marco~Patricio
  Crasso}, \bibinfo{person}{Dakuo Wang}, \bibinfo{person}{Michael Muller},
  \bibinfo{person}{N~Sadat Shami}, {and} \bibinfo{person}{Werner Geyer}.}
  \bibinfo{year}{2018}\natexlab{}.
\newblock \showarticletitle{All Work and No Play? Conversations With a
  Question-And-Answer Chatbot in the Wild}. In
  \bibinfo{booktitle}{\emph{Proceedings of the 2018 CHI Conference on Human
  Factors in Computing Systems}}. \bibinfo{publisher}{ACM},
  \bibinfo{pages}{1--13}.
\newblock
\urldef\tempurl%
\url{https://doi.org/10.1145/3173574.3173577}
\showDOI{\tempurl}


\bibitem[\protect\citeauthoryear{Liu and Heer}{Liu and Heer}{2014}]%
        {latency}
\bibfield{author}{\bibinfo{person}{Zhicheng Liu} {and} \bibinfo{person}{Jeffrey
  Heer}.} \bibinfo{year}{2014}\natexlab{}.
\newblock \showarticletitle{The Effects of Interactive Latency on Exploratory
  Visual Analysis}.
\newblock \bibinfo{journal}{\emph{IEEE Transactions on Visualization and
  Computer Graphics}} \bibinfo{volume}{20}, \bibinfo{number}{12}
  (\bibinfo{year}{2014}), \bibinfo{pages}{2122--2131}.
\newblock
\urldef\tempurl%
\url{https://doi.org/10.1109/TVCG.2014.2346452}
\showDOI{\tempurl}


\bibitem[\protect\citeauthoryear{m00nlight Wang}{m00nlight Wang}{2018}]%
        {medium2}
\bibfield{author}{\bibinfo{person}{m00nlight Wang}.}
  \bibinfo{year}{2018}\natexlab{}.
\newblock \bibinfo{booktitle}{\emph{Income Inequality Analysis and
  Visualization}}.
\newblock
\urldef\tempurl%
\url{https://medium.com/@m00nlight/income-inequality-analysis-and-visualization-f688a4fc6609}
\showURL{%
\tempurl}


\bibitem[\protect\citeauthoryear{Matute, Blanco, Yarritu, D{\'\i}az-Lago,
  Vadillo, and Barberia}{Matute et~al\mbox{.}}{2015}]%
        {causalIllusion}
\bibfield{author}{\bibinfo{person}{Helena Matute}, \bibinfo{person}{Fernando
  Blanco}, \bibinfo{person}{Ion Yarritu}, \bibinfo{person}{Marcos
  D{\'\i}az-Lago}, \bibinfo{person}{Miguel~A Vadillo}, {and}
  \bibinfo{person}{Itxaso Barberia}.} \bibinfo{year}{2015}\natexlab{}.
\newblock \showarticletitle{Illusions of Causality: How They Bias Our Everyday
  Thinking and How They Could Be Reduced}.
\newblock \bibinfo{journal}{\emph{Frontiers in Psychology}}
  \bibinfo{volume}{6} (\bibinfo{year}{2015}), \bibinfo{pages}{888}.
\newblock
\urldef\tempurl%
\url{https://doi.org/10.3389/fpsyg.2015.00888}
\showDOI{\tempurl}


\bibitem[\protect\citeauthoryear{Micallef, Palmas, Oulasvirta, and
  Weinkauf}{Micallef et~al\mbox{.}}{2017}]%
        {scatterplotOp}
\bibfield{author}{\bibinfo{person}{Luana Micallef}, \bibinfo{person}{Gregorio
  Palmas}, \bibinfo{person}{Antti Oulasvirta}, {and} \bibinfo{person}{Tino
  Weinkauf}.} \bibinfo{year}{2017}\natexlab{}.
\newblock \showarticletitle{Towards Perceptual Optimization of the Visual
  Design of Scatterplots}.
\newblock \bibinfo{journal}{\emph{IEEE Transactions on Visualization and
  Computer Graphics}} \bibinfo{volume}{23}, \bibinfo{number}{6}
  (\bibinfo{year}{2017}), \bibinfo{pages}{1588--1599}.
\newblock
\urldef\tempurl%
\url{https://doi.org/10.1109/TVCG.2017.2674978}
\showDOI{\tempurl}


\bibitem[\protect\citeauthoryear{Moere, Tomitsch, Wimmer, Christoph, and
  Grechenig}{Moere et~al\mbox{.}}{2012}]%
        {styles}
\bibfield{author}{\bibinfo{person}{Andrew~Vande Moere}, \bibinfo{person}{Martin
  Tomitsch}, \bibinfo{person}{Christoph Wimmer}, \bibinfo{person}{Boesch
  Christoph}, {and} \bibinfo{person}{Thomas Grechenig}.}
  \bibinfo{year}{2012}\natexlab{}.
\newblock \showarticletitle{Evaluating the Effect of Style in Information
  Visualization}.
\newblock \bibinfo{journal}{\emph{IEEE Transactions on Visualization and
  Computer Graphics}} \bibinfo{volume}{18}, \bibinfo{number}{12}
  (\bibinfo{year}{2012}), \bibinfo{pages}{2739--2748}.
\newblock
\urldef\tempurl%
\url{https://doi.org/10.1109/TVCG.2012.221}
\showDOI{\tempurl}


\bibitem[\protect\citeauthoryear{Monday}{Monday}{2020}]%
        {makeover}
\bibfield{author}{\bibinfo{person}{Makeover Monday}.}
  \bibinfo{year}{2020}\natexlab{}.
\newblock \bibinfo{title}{Makeover Monday $|$ a Weekly Social Data Project}.
\newblock
\newblock
\urldef\tempurl%
\url{https://www.makeovermonday.co.uk}
\showURL{%
\tempurl}


\bibitem[\protect\citeauthoryear{Pandey, Manivannan, Nov, Satterthwaite, and
  Bertini}{Pandey et~al\mbox{.}}{2014}]%
        {persuasive}
\bibfield{author}{\bibinfo{person}{Anshul~Vikram Pandey},
  \bibinfo{person}{Anjali Manivannan}, \bibinfo{person}{Oded Nov},
  \bibinfo{person}{Margaret Satterthwaite}, {and} \bibinfo{person}{Enrico
  Bertini}.} \bibinfo{year}{2014}\natexlab{}.
\newblock \showarticletitle{The Persuasive Power of Data Visualization}.
\newblock \bibinfo{journal}{\emph{IEEE Transactions on Visualization and
  Computer Graphics}} \bibinfo{volume}{20}, \bibinfo{number}{12}
  (\bibinfo{year}{2014}), \bibinfo{pages}{2211--2220}.
\newblock
\urldef\tempurl%
\url{https://doi.org/10.1109/TVCG.2014.2346419}
\showDOI{\tempurl}


\bibitem[\protect\citeauthoryear{Pandey, Rall, Satterthwaite, Nov, and
  Bertini}{Pandey et~al\mbox{.}}{2015}]%
        {deceptive}
\bibfield{author}{\bibinfo{person}{Anshul~Vikram Pandey},
  \bibinfo{person}{Katharina Rall}, \bibinfo{person}{Margaret~L Satterthwaite},
  \bibinfo{person}{Oded Nov}, {and} \bibinfo{person}{Enrico Bertini}.}
  \bibinfo{year}{2015}\natexlab{}.
\newblock \showarticletitle{How Deceptive Are Deceptive Visualizations? an
  Empirical Analysis of Common Distortion Techniques}. In
  \bibinfo{booktitle}{\emph{Proceedings of the 2015 CHI Conference on Human
  Factors in Computing Systems}}. \bibinfo{publisher}{ACM},
  \bibinfo{pages}{1469--1478}.
\newblock
\urldef\tempurl%
\url{https://doi.org/10.1145/2702123.2702608}
\showDOI{\tempurl}


\bibitem[\protect\citeauthoryear{Pu, Chen, and Hu}{Pu et~al\mbox{.}}{2011}]%
        {recommender}
\bibfield{author}{\bibinfo{person}{Pearl Pu}, \bibinfo{person}{Li Chen}, {and}
  \bibinfo{person}{Rong Hu}.} \bibinfo{year}{2011}\natexlab{}.
\newblock \showarticletitle{A User-Centric Evaluation Framework for Recommender
  System}. In \bibinfo{booktitle}{\emph{Proceedings of the Fifth ACM Conference
  on Recommender Systems}}. \bibinfo{publisher}{ACM},
  \bibinfo{pages}{157--164}.
\newblock
\urldef\tempurl%
\url{https://doi.org/10.1145/2043932.2043962}
\showDOI{\tempurl}


\bibitem[\protect\citeauthoryear{Ritchie, Wigdor, and Chevalier}{Ritchie
  et~al\mbox{.}}{2019}]%
        {lie}
\bibfield{author}{\bibinfo{person}{Jacob Ritchie}, \bibinfo{person}{Daniel
  Wigdor}, {and} \bibinfo{person}{Fann Chevalier}.}
  \bibinfo{year}{2019}\natexlab{}.
\newblock \showarticletitle{A Lie Reveals the Truth: Quasimodes for
  Task-Aligned Data Presentation}. In \bibinfo{booktitle}{\emph{Proceedings of
  the 2019 CHI Conference on Human Factors in Computing Systems}}.
  \bibinfo{publisher}{ACM}, \bibinfo{pages}{1--13}.
\newblock
\urldef\tempurl%
\url{https://doi.org/10.1145/3290605.3300423}
\showDOI{\tempurl}


\bibitem[\protect\citeauthoryear{Rohrer}{Rohrer}{2018}]%
        {causalInference}
\bibfield{author}{\bibinfo{person}{Julia~M Rohrer}.}
  \bibinfo{year}{2018}\natexlab{}.
\newblock \showarticletitle{Thinking Clearly About Correlations and Causation:
  Graphical Causal Models for Observational Data}.
\newblock \bibinfo{journal}{\emph{Advances in Methods and Practices in
  Psychological Science}} \bibinfo{volume}{1}, \bibinfo{number}{1}
  (\bibinfo{year}{2018}), \bibinfo{pages}{27--42}.
\newblock
\urldef\tempurl%
\url{https://doi.org/10.1177/2515245917745629}
\showDOI{\tempurl}


\bibitem[\protect\citeauthoryear{Sacha, Stoffel, Stoffel, Kwon, Ellis, and
  Keim}{Sacha et~al\mbox{.}}{2014}]%
        {knowledge}
\bibfield{author}{\bibinfo{person}{Dominik Sacha}, \bibinfo{person}{Andreas
  Stoffel}, \bibinfo{person}{Florian Stoffel}, \bibinfo{person}{Bum~Chul Kwon},
  \bibinfo{person}{Geoffrey Ellis}, {and} \bibinfo{person}{Daniel~A Keim}.}
  \bibinfo{year}{2014}\natexlab{}.
\newblock \showarticletitle{Knowledge Generation Model for Visual Analytics}.
\newblock \bibinfo{journal}{\emph{IEEE Transactions on Visualization and
  Computer Graphics}} \bibinfo{volume}{20}, \bibinfo{number}{12}
  (\bibinfo{year}{2014}), \bibinfo{pages}{1604--1613}.
\newblock
\urldef\tempurl%
\url{https://doi.org/10.1109/TVCG.2014.2346481}
\showDOI{\tempurl}


\bibitem[\protect\citeauthoryear{Sarikaya and Gleicher}{Sarikaya and
  Gleicher}{2017}]%
        {scatterDesign}
\bibfield{author}{\bibinfo{person}{Alper Sarikaya} {and}
  \bibinfo{person}{Michael Gleicher}.} \bibinfo{year}{2017}\natexlab{}.
\newblock \showarticletitle{Scatterplots: Tasks, Data, and Designs}.
\newblock \bibinfo{journal}{\emph{IEEE Transactions on Visualization and
  Computer Graphics}} \bibinfo{volume}{24}, \bibinfo{number}{1}
  (\bibinfo{year}{2017}), \bibinfo{pages}{402--412}.
\newblock
\urldef\tempurl%
\url{https://doi.org/10.1109/TVCG.2017.2744184}
\showDOI{\tempurl}


\bibitem[\protect\citeauthoryear{Setlur, Battersby, Tory, Gossweiler, and
  Chang}{Setlur et~al\mbox{.}}{2016}]%
        {eviza}
\bibfield{author}{\bibinfo{person}{Vidya Setlur}, \bibinfo{person}{Sarah~E
  Battersby}, \bibinfo{person}{Melanie Tory}, \bibinfo{person}{Rich
  Gossweiler}, {and} \bibinfo{person}{Angel~X Chang}.}
  \bibinfo{year}{2016}\natexlab{}.
\newblock \showarticletitle{Eviza: A Natural Language Interface for Visual
  Analysis}. In \bibinfo{booktitle}{\emph{Proceedings of the 29th Annual ACM
  Symposium on User Interface Software \& Technology}}.
  \bibinfo{publisher}{ACM}, \bibinfo{pages}{365--377}.
\newblock
\urldef\tempurl%
\url{https://doi.org/10.1145/2984511.2984588}
\showDOI{\tempurl}


\bibitem[\protect\citeauthoryear{Sinha and Swearingen}{Sinha and
  Swearingen}{2002}]%
        {transparency}
\bibfield{author}{\bibinfo{person}{Rashmi Sinha} {and} \bibinfo{person}{Kirsten
  Swearingen}.} \bibinfo{year}{2002}\natexlab{}.
\newblock \showarticletitle{The Role of Transparency in Recommender Systems}.
  In \bibinfo{booktitle}{\emph{Extended Abstracts of the 2002 CHI Conference on
  Human Factors in Computing Systems}}. \bibinfo{publisher}{ACM},
  \bibinfo{pages}{830--831}.
\newblock
\urldef\tempurl%
\url{https://doi.org/10.1145/506443.506619}
\showDOI{\tempurl}


\bibitem[\protect\citeauthoryear{Srinivasan, Drucker, Endert, and
  Stasko}{Srinivasan et~al\mbox{.}}{2018}]%
        {voder}
\bibfield{author}{\bibinfo{person}{Arjun Srinivasan}, \bibinfo{person}{Steven~M
  Drucker}, \bibinfo{person}{Alex Endert}, {and} \bibinfo{person}{John
  Stasko}.} \bibinfo{year}{2018}\natexlab{}.
\newblock \showarticletitle{Augmenting Visualizations With Interactive Data
  Facts to Facilitate Interpretation and Communication}.
\newblock \bibinfo{journal}{\emph{IEEE Transactions on Visualization and
  Computer Graphics}} \bibinfo{volume}{25}, \bibinfo{number}{1}
  (\bibinfo{year}{2018}), \bibinfo{pages}{672--681}.
\newblock
\urldef\tempurl%
\url{https://doi.org/10.1109/TVCG.2018.2865145}
\showDOI{\tempurl}


\bibitem[\protect\citeauthoryear{Srinivasan and Stasko}{Srinivasan and
  Stasko}{2017}]%
        {orko}
\bibfield{author}{\bibinfo{person}{Arjun Srinivasan} {and}
  \bibinfo{person}{John Stasko}.} \bibinfo{year}{2017}\natexlab{}.
\newblock \showarticletitle{Orko: Facilitating Multimodal Interaction for
  Visual Exploration and Analysis of Networks}.
\newblock \bibinfo{journal}{\emph{IEEE Transactions on Visualization and
  Computer Graphics}} \bibinfo{volume}{24}, \bibinfo{number}{1}
  (\bibinfo{year}{2017}), \bibinfo{pages}{511--521}.
\newblock
\urldef\tempurl%
\url{https://doi.org/10.1109/TVCG.2017.2745219}
\showDOI{\tempurl}


\bibitem[\protect\citeauthoryear{Tableau}{Tableau}{2019a}]%
        {tTutorial1}
\bibfield{author}{\bibinfo{person}{Tableau}.} \bibinfo{year}{2019}\natexlab{a}.
\newblock \bibinfo{title}{{E}xplain {D}ata Internals: Automated Bayesian
  Modeling $|$ {T}ableau {C}onference 2019}.
\newblock
\newblock
\urldef\tempurl%
\url{https://tc19.tableau.com/learn/sessions/explain-data-internals-automated-bayesian-modeling?_ga=2.242994050.1845292459.1583776901-580893601.1583776901&_fsi=H59ZIxRV}
\showURL{%
\tempurl}


\bibitem[\protect\citeauthoryear{Tableau}{Tableau}{2019b}]%
        {tTutorial2}
\bibfield{author}{\bibinfo{person}{Tableau}.} \bibinfo{year}{2019}\natexlab{b}.
\newblock \bibinfo{title}{Inspect a View using {E}xplain {D}ata -- {T}ableau}.
\newblock
\newblock
\urldef\tempurl%
\url{https://help.tableau.com/current/pro/desktop/en-us/explain_data.htm}
\showURL{%
\tempurl}


\bibitem[\protect\citeauthoryear{Tableau}{Tableau}{2020}]%
        {explainData}
\bibfield{author}{\bibinfo{person}{Tableau}.} \bibinfo{year}{2020}\natexlab{}.
\newblock \bibinfo{title}{Explain Data | Tableau Software}.
\newblock
\newblock
\urldef\tempurl%
\url{https://www.tableau.com/products/new-features/explain-data}
\showURL{%
\tempurl}


\bibitem[\protect\citeauthoryear{Valdez, Ziefle, and Sedlmair}{Valdez
  et~al\mbox{.}}{2017}]%
        {priming}
\bibfield{author}{\bibinfo{person}{Andre~Calero Valdez},
  \bibinfo{person}{Martina Ziefle}, {and} \bibinfo{person}{Michael Sedlmair}.}
  \bibinfo{year}{2017}\natexlab{}.
\newblock \showarticletitle{Priming and Anchoring Effects in Visualization}.
\newblock \bibinfo{journal}{\emph{IEEE Transactions on Visualization and
  Computer Graphics}} \bibinfo{volume}{24}, \bibinfo{number}{1}
  (\bibinfo{year}{2017}), \bibinfo{pages}{584--594}.
\newblock
\urldef\tempurl%
\url{https://doi.org/10.1109/TVCG.2017.2744138}
\showDOI{\tempurl}


\bibitem[\protect\citeauthoryear{Wang and Mueller}{Wang and Mueller}{2015}]%
        {cGraph}
\bibfield{author}{\bibinfo{person}{Jun Wang} {and} \bibinfo{person}{Klaus
  Mueller}.} \bibinfo{year}{2015}\natexlab{}.
\newblock \showarticletitle{The Visual Causality Analyst: An Interactive
  Interface for Causal Reasoning}.
\newblock \bibinfo{journal}{\emph{IEEE Transactions on Visualization and
  Computer Graphics}} \bibinfo{volume}{22}, \bibinfo{number}{1}
  (\bibinfo{year}{2015}), \bibinfo{pages}{230--239}.
\newblock
\urldef\tempurl%
\url{https://doi.org/10.1109/TVCG.2015.2467931}
\showDOI{\tempurl}


\bibitem[\protect\citeauthoryear{Wogalter}{Wogalter}{2006}]%
        {model}
\bibfield{author}{\bibinfo{person}{Michael~S Wogalter}.}
  \bibinfo{year}{2006}\natexlab{}.
\newblock \showarticletitle{Communication-Human Information Processing (C-HIP)
  Model}.
\newblock \bibinfo{journal}{\emph{Handbook of warnings}}
  (\bibinfo{year}{2006}), \bibinfo{pages}{51--61}.
\newblock


\bibitem[\protect\citeauthoryear{Xiong, Shapiro, Hullman, and Franconeri}{Xiong
  et~al\mbox{.}}{2019a}]%
        {correlation}
\bibfield{author}{\bibinfo{person}{Cindy Xiong}, \bibinfo{person}{Joel
  Shapiro}, \bibinfo{person}{Jessica Hullman}, {and} \bibinfo{person}{Steven
  Franconeri}.} \bibinfo{year}{2019}\natexlab{a}.
\newblock \showarticletitle{Illusion of Causality in Visualized Data}.
\newblock \bibinfo{journal}{\emph{IEEE Transactions on Visualization and
  Computer Graphics}} \bibinfo{volume}{26}, \bibinfo{number}{1}
  (\bibinfo{year}{2019}), \bibinfo{pages}{853--862}.
\newblock
\urldef\tempurl%
\url{https://doi.org/10.1109/TVCG.2019.2934399}
\showDOI{\tempurl}


\bibitem[\protect\citeauthoryear{Xiong, van Weelden, and Franconeri}{Xiong
  et~al\mbox{.}}{2019b}]%
        {curse}
\bibfield{author}{\bibinfo{person}{Cindy Xiong}, \bibinfo{person}{Lisanne van
  Weelden}, {and} \bibinfo{person}{Steven Franconeri}.}
  \bibinfo{year}{2019}\natexlab{b}.
\newblock \showarticletitle{The Curse of Knowledge in Visual Data
  Communication}.
\newblock \bibinfo{journal}{\emph{IEEE Transactions on Visualization and
  Computer Graphics}} (\bibinfo{year}{2019}).
\newblock
\urldef\tempurl%
\url{https://doi.org/10.1109/TVCG.2019.2917689}
\showDOI{\tempurl}


\bibitem[\protect\citeauthoryear{Yu and Silva}{Yu and Silva}{2019}]%
        {flowsense}
\bibfield{author}{\bibinfo{person}{Bowen Yu} {and}
  \bibinfo{person}{Cl{\'a}udio~T Silva}.} \bibinfo{year}{2019}\natexlab{}.
\newblock \showarticletitle{Flowsense: A Natural Language Interface for Visual
  Data Exploration Within a Dataflow System}.
\newblock \bibinfo{journal}{\emph{IEEE Transactions on Visualization and
  Computer Graphics}} \bibinfo{volume}{26}, \bibinfo{number}{1}
  (\bibinfo{year}{2019}), \bibinfo{pages}{1--11}.
\newblock
\urldef\tempurl%
\url{https://doi.org/10.1109/TVCG.2019.2934668}
\showDOI{\tempurl}


\bibitem[\protect\citeauthoryear{Zhao, Qu, and Sedlmair}{Zhao
  et~al\mbox{.}}{2019}]%
        {neighborhood}
\bibfield{author}{\bibinfo{person}{Mingqian Zhao}, \bibinfo{person}{Huamin Qu},
  {and} \bibinfo{person}{Michael Sedlmair}.} \bibinfo{year}{2019}\natexlab{}.
\newblock \showarticletitle{Neighborhood Perception in Bar Charts}. In
  \bibinfo{booktitle}{\emph{Proceedings of the 2019 CHI Conference on Human
  Factors in Computing Systems}}. \bibinfo{publisher}{ACM},
  \bibinfo{pages}{1--12}.
\newblock
\urldef\tempurl%
\url{https://doi.org/10.1145/3290605.3300462}
\showDOI{\tempurl}


\bibitem[\protect\citeauthoryear{Zhi and Metoyer}{Zhi and Metoyer}{2020}]%
        {sports}
\bibfield{author}{\bibinfo{person}{Qiyu Zhi} {and} \bibinfo{person}{Ronald
  Metoyer}.} \bibinfo{year}{2020}\natexlab{}.
\newblock \showarticletitle{GameBot: A Visualization-Augmented Chatbot for
  Sports Game}. In \bibinfo{booktitle}{\emph{Extended Abstracts of the 2020 CHI
  Conference on Human Factors in Computing Systems}}. \bibinfo{publisher}{ACM},
  \bibinfo{pages}{1--7}.
\newblock
\urldef\tempurl%
\url{https://doi.org/10.1145/3334480.3382794}
\showDOI{\tempurl}


\end{thebibliography}

\end{document}